\documentclass[aps,prx,twocolumn,showpacs,amsmath,amssymb,superscriptaddress,floatfix,reprint,nofootinbib]{revtex4-2}
\usepackage{amssymb}
\usepackage{amsbsy}
\usepackage{amsmath}
\usepackage{mathtools}
\usepackage{mathrsfs}
\usepackage{epsfig}
\usepackage{graphicx}

\usepackage{array}
\usepackage{textcomp}
\usepackage{xcolor}
\usepackage{braket}
\usepackage{hhline}
\usepackage{dcolumn}
\usepackage{dsfont}
\usepackage{comment}
\usepackage{bm,hyperref}
\usepackage[justification=raggedright,font=small]{caption}
\usepackage[titletoc,toc,title]{appendix}
\usepackage[normalem]{ulem}
\usepackage[caption=false]{subfig}
\usepackage{cleveref}
\usepackage{multirow}
\usepackage{algpseudocode}
\usepackage{algorithm}

\crefname{equation}{Eq.}{Eqs.}
\hypersetup{
  pdffitwindow=false,            
  pdfstartview={Fit},            
  colorlinks=true,               
  linkcolor=blue,      
  citecolor=blue,       
  urlcolor=magenta           
}
\newcommand{\tr}{\text{Tr}}
\newcommand{\eq}[2]{\begin{equation} \label{#1} #2 \end{equation}}
\newcommand{\eqsp}[1]{\begin{split} #1 \end{split}}

\newcommand{\bit}{\begin{itemize}}
\newcommand{\eit}{\end{itemize}}

\newcommand{\be}{\begin{equation}}
\newcommand{\ee}{\end{equation}}
\newcommand{\iea}{\begin{equation}\begin{aligned}}
\newcommand{\fea}{\end{aligned}\end{equation}}

\newcommand{\mbZ}{\mathbb{Z}}

\begin{document}

\author{Sanket Chirame}
\affiliation{School of Physics and Astronomy, University of Minnesota, Minneapolis, Minnesota 55455, USA}
\author{Abhinav Prem}
\affiliation{School of Natural Sciences, Institute for Advanced Study, Princeton, New Jersey 08540, USA}
\affiliation{Physics Program, Bard College, 30 Campus Road, Annandale-on-Hudson, New York 12504, USA}
\author{Sarang Gopalakrishnan}
\affiliation{Department of Electrical and Computer Engineering, Princeton University, Princeton, NJ 08544}
\author{Fiona J. Burnell}
\affiliation{School of Physics and Astronomy, University of Minnesota, Minneapolis, Minnesota 55455, USA}

\title{Stabilizing Non-Abelian Topological Order against Heralded Noise via Local Lindbladian Dynamics}
\date{\today}
\begin{abstract}

An important open question for the current generation of highly controllable quantum devices is understanding which phases can be realized as stable steady-states under local quantum dynamics. In this work, we show how robust steady-state phases with both Abelian and non-Abelian mixed-state topological order can be stabilized, in two spatial dimensions (2d), against generic ``heralded" noise using active dynamics that incorporate measurement and feedback, modeled as a \emph{fully local} Lindblad master equation. These topologically ordered steady states are two-way connected to pure topologically ordered ground states using local quantum channels, and preserve quantum information for a time that is exponentially large in the system size. Specifically, we present explicit constructions of families of local Lindbladians for both Abelian ($\mathbb{Z}_2$) and non-Abelian ($D_4$) topological order whose steady-states host mixed-state topological order when the noise is below a threshold strength. As the noise strength is increased, these models exhibit first-order transitions to intermediate mixed state phases where they encode robust classical memories, followed by (first-order) transitions to a trivial steady state at high noise rates. When the noise is imperfectly heralded, steady-state order disappears but our active dynamics significantly enhances the lifetime of the encoded logical information. To carry out the numerical simulations for the non-Abelian $D_4$ case, we introduce a generalized stabilizer tableau formalism that permits efficient simulation of the non-Abelian Lindbladian dynamics.

\end{abstract}
\maketitle



\section{Introduction}
\label{sec:intro}

Topological phases of matter are characterized by a number of intriguing properties, including multiple locally indistinguishable ground states with subtle entanglement structure that enables them to store quantum information.  However, while these phases are stable to local perturbations of the Hamiltonian that do not close the many-body gap~\cite{bravyi2010,bravyi2011short,michalakis2013}, in $d \leq 3$, topological order (TO) does not survive at any nonzero temperature~\cite{hastings2011finiteT,bravyinogo,poulin2013,terhalreview,brownreview,roberts2017finiteT,lu2020negativity,stahl2021finiteT}.\footnote{This is fundamentally due to the presence of deconfined point-like excitations which, when thermally excited, destroy the long-range entanglement in the ground-space (even if the excitations have restricted mobility, as in fracton models~\cite{chamon2005,kim2016,prem2017}). However, certain models in $d \geq 4$ (e.g., the 4d Toric code~\cite{dennis2002,alicki2010}) lack any point-like excitations and are thermally stable phases~\cite{bombin2013scm,brownreview}.}

The fact that even a very cold environment is incompatible with low-dimensional TO raises the question of whether such orders can be stabilized at all in noisy open quantum systems.  
One approach to achieving this in two spatial dimensions (2d) is quantum error correction~\cite{dennis2002,kay2008,wang2009,woottonreview,wootton2014nonabelianQEC,brell2014,wootton2016,DauphinaisPoulin,schotte2023,lihm2018implementation}, which, when performed at regular intervals, can protect information stored in the ground space of a topologically ordered Hamiltonian (which we henceforth refer to as a TO state) for arbitrarily long times.  
More specifically, in a TO state that is subjected to sufficiently weak noise for a sufficiently short time,  a round of error-syndrome measurements, followed by a (typically nonlocal) recovery operation, is sufficient to fully recover the original state.   
Conversely, if the noise is too strong or acts for too long time, information about the initial state is irretrievably lost in an information-theoretic phase transition.  
The existence of such mixed-state phase transitions has led to a surge of interest in investigating topologically non-trivial mixed states that result from local noise channels~\cite{coser2019class,degroot2022og,molnar2022mpo1,molnar2022mpo2,lee2023,ma2023prx,bao2023mixed,fan2023mixed,zhang2022strange,lee2022aspt,chen2023separable,chen2023separable2,guo2023cluster2d,paszko2023,sohal2024imto,ellison2024,li2024,chirame2024spt,alexaspt,sala2024,lu2024dis,lessa2024ssb,sala2024D4decoherence,sala2024stabilityloopmodel,shah2024instability} and in defining notions of mixed-state phases of matter~\cite{ma2023avg,rakovszky2023stable,lessa2024anomaly,sang2023mixed,sang2024,zhang2024,sang2024def,thompson2024population}.

Mixed-state phases are often studied in the context of finite-time dissipative evolution; indeed, the error-correcting transition is a transition in recoverability that occurs \emph{at finite time}.  However, the most natural out of equilibrium analogs of equilibrium phases of matter, which are by definition stationary in time, are \emph{steady-states}.  
Clearly, one route to achieving steady-state TO is through repeated rounds of weak noise and error correction.
However, because quantum error correction involves non-local recovery processes, a dynamical regime that can be stabilized only by error correction differs sharply from a phase of matter, for which spatial locality and the Lieb-Robinson bound are crucial defining properties.  
To have a well-defined \textit{steady-state phase}, we must require that the full quantum dynamics, including measurements, propagation of classical information, and the resulting feedback operations,  is fully local. Against a general noise model, it is not clear that any such topologically ordered phases exist in $d \leq 3$.

The central result of this paper is that a physically motivated restriction of the noise model \emph{does} allow for a local error-correction procedure with a nonzero threshold, and therefore to a nontrivial steady-state phase that protects quantum information. Specifically, we consider noise that is \emph{heralded}, in the sense that one knows which sites are potentially corrupted by noise. Because heralded noise (also called erasure noise) is easier to correct than generic noise, many present-day experimental platforms rely on ``erasure conversion'' techniques to herald noise events; almost perfect heralding can be achieved with neutral atoms~\cite{grassl1997,knill2003,wu2022,kubica2023,sahay2023,gu2023erasure,gu2024}. Remarkably, heralded noise is much easier to correct locally than generic noise. We showed previously in Ref.~\cite{chirame2024spt} that in 1d systems with biased erasure noise, one can stabilize symmetry protected topological (SPT) order for times that are exponentially long in system size. Here, we show that in 2d one can achieve much more while requiring less---for \emph{any} heralded noise, one can construct local error-correction protocols with nonzero thresholds that stabilize both the Abelian Toric code TO and a non-Abelian TO associated with the group $D_4$, for times that scale exponentially in the (linear) system size. Thus, we demonstrate the existence of steady-state phases with mixed-state non-Abelian TO that are stable against general heralded noise.

The rest of this paper is structured as follows: in Sec.~\ref{sec:background}, we introduce the key background concepts we use and summarize our main results. In Sec.~\ref{sec:TC}, we first illustrate our local error-correction protocol for the simpler case of the 2d Toric code, which hosts Abelian TO. We detail the heralded noise model, the correction protocol, and define the pertinent metrics we use to establish the existence of a topologically ordered steady-state phase. In Sec.~\ref{sec:D4}, we provide an overview of the non-Abelian $D_4$ lattice model considered here (recently prepared experimentally by Ref.~\cite{iqbal2024nonabelian}) and discuss our local correction protocol. Here, we also introduce the stabilizer tableau method used to simulate the full Lindbladian dynamics in the non-Abelian system. Then, in Sec.~\ref{sec:phases} we present the results of our simulations, including the steady-state phase diagram of the $D_4$ model, a characterization of the transitions between distinct dynamical phases, and the effect of imperfectly heralded errors. We conclude with a discussion of open questions in Sec.~\ref{sec:cncls}.


\section{Background and Summary of Main Results}
\label{sec:background}

\begin{figure}[t]
\includegraphics{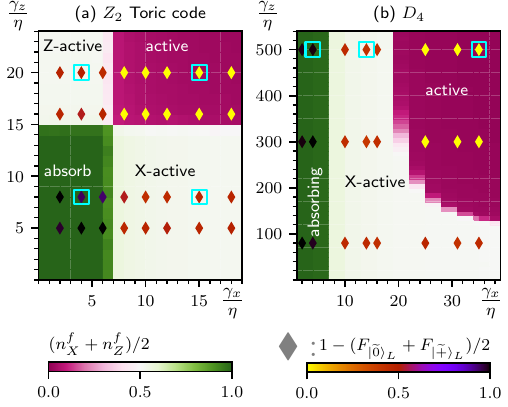}
\caption{\label{fig:2dphase}Steady-state phase diagrams for (a) the Toric code on a $48\times 48 $ honeycomb lattice and (b) the non-Abelian $D_4$ model defined on three interpenetrating $48\times 48$ sized honeycomb lattices. The active dynamics removes the $X$ and $Z$ flags at rates $\gamma_x$ and $\gamma_z$, respectively. The heralded depolarizing noise rate is set to $\eta=1$. The background color indicates the combined density $\frac{1}{2}(n^f_X+n^f_Z)$ of flags evaluated at a late time ($t_f=100$ for the Toric code and $t_f=20$ for the $D_4$ model) at which the densities (for parameters away from the phase boundaries) have saturated to time-independent values (see Appendix~\ref{app:numerics}). The logical error rates are computed for a select set of parameters depicted using the diamond-shaped symbols. These are colored according to $1-\frac{1}{2}(F_{|\widetilde{0}\rangle_L}+F_{|\widetilde{+}\rangle_L})$, which quantifies the steady-state logical error rate averaged over two distinct initial logical states. Each data point is obtained by averaging over $100$ independent Monte Carlo realizations. 
For both models, the active phase of the flags ($n^f_X,n^f_Z\approx 0$) leads to a topologically ordered steady-state phase, where both logical fidelities $F$ are close to $1$. This phase protects the initially encoded quantum information up to exponentially long times. In contrast, the partially active phases (only one type of flags has small density) can only protect classical information for long times, signaled by only one of the $F\approx 1$. The detailed time evolution for representative points in each of these phases, highlighted using cyan colored squares, is shown in Fig.~\ref{fig:TC-MC-dynamics} and \ref{fig:d4-densityVst}.} 
\end{figure}

\subsection*{Background}

Before turning to our central results, we briefly review the concepts of heralded noise and steady-state phases. Typical qubits in NISQ devices are subjected to errors that are undetectable within the qubit subspace; in contrast, in certain settings the dominant error processes correspond to leakage errors (also called ``erasure" or ``flagged" errors) out of the computational subspace. These errors are then experimentally detected in real time by measurements which do not disturb  the state of qubits within the logical space;  qubits for which erasures are detected are then reinitialized in an arbitrary initial state, which therefore has some probability of introducing an error.  Importantly, however, the location of these erasures can be stored in a classical register, giving information about the possible error locations. Systems where heralded errors are the dominant noise source include Rydberg atom arrays with biased erasure noise~\cite{wu2022,sahay2023,maHighfidelityGatesMidcircuit2023,scholl2023erasure}, trapped ions~\cite{campbell2020,kang2023}, and dual-rail superconducting cavity qubits with erasure-converted errors~\cite{kubica2023,teoh2023,chou2024,kootandavida2024,levine2024}. 

In the context of hardware-efficient QEC, systems where erasure errors are the dominant noise sources are being actively studied given that such errors are more efficiently correctable, leading to higher thresholds and reduced overhead~\cite{grassl1997,knill2003,wu2022,kubica2023,sahay2023,gu2023erasure,gu2024}. Prior work has investigated the performance of classical decoding algorithms for surface codes~\cite{KitaevToric} in the presence of erasure errors; for instance, Ref.~\cite{stace2009} found that error correction fails when the percolated clusters of erased qubits span the lattice. More recently, Ref.~\cite{delfosse2020erasure} developed a decoder for surface codes assuming perfectly heralded errors: for a given erasure pattern, all loops on the square lattice are first removed while ensuring that the erasures still span all syndromes, which are then efficiently paired up. The Union-Find decoder~\cite{delfosse2021almostlineartime}, an efficient and widely-used decoder, extends this protocol to include the effect of unheralded Pauli errors. In either case, the protocols are non-local.  

Here, we show that heralding can also be exploited to develop a fully \textit{local} correction protocol, which stabilizes steady-state TO in the presence of continuous heralded noise.  A key requirement for success is the ability to collapse {\it loops} of erasures using purely local processes, whilst maintaining perfect heralding of the syndromes.  Our protocol does this by dynamically deforming corners of the erasure pattern in a directionally biased way. We detail below how this can be done in such a way that all syndromes lie along a string of flags, thereby maintaining perfect heralding and allowing the continuously acting correction protocol to stabilize a topological steady-state.  Locality may offer practical benefits for error-correction in terms of scalability; however, our main interest here is that such local dynamics allows us to define physically meaningful equivalence relations on the space of steady-states that arise under local Markovian evolution-- i.e., to define distinct robust, mixed-state steady-state phases~\cite{coser2019class,lessa2024anomaly,sang2023mixed,rakovszky2023stable,sang2024}. 

The protocol we consider can be written as a Lindblad master equation $\partial_t \rho = \mathcal{L}(\rho)$ with two types of jump operators, corresponding respectively to the heralded noise and the recovery procedure. We have specified the recovery procedure as a ``measurement-and-feedback'' process; however, any jump operator can be written in this form through a polar decomposition. We are interested in the steady-state manifolds of the Lindblad superoperator $\mathcal{L}$ as a function of the noise and recovery rates. Steady-states are eigenvectors of $\mathcal{L}$ with zero eigenvalue. In a stable TO phase, we expect $\mathcal{L}$ to have a nontrivial manifold of states with eigenvalues that are exponentially close in modulus to $0$ (in linear system size).

We will primarily define steady-state phases via the ``uniformity'' criterion~\cite{rakovszky2023stable}. 
This criterion can be stated informally as follows. Given two Lindbladians $\mathcal{L}, \mathcal{L}'$, suppose every steady-state of $\mathcal{L}$ can be quickly reached by starting with a steady-state of $\mathcal{L}'$ and evolving for a short time, and vice versa. Then we say that $\mathcal{L}$ and $\mathcal{L}'$ (or equivalently their steady-state manifolds) are in the same phase\footnote{In fact, it is sufficient for the steady-states of $\mathcal{L}, \mathcal{L}'$ to be connected by a short-time evolution involving a time-dependent Lindbladian that interpolates between the two.}. We will test this criterion explicitly, starting with the steady-state of the noisy evolution, turning off the noise, and checking that we rapidly approach a steady-state of the noiseless evolution. Note that when this relaxation time is at most $\mathcal{O}(\text{polylog}(L))$, this criterion is equivalent to establishing two-way channel connectivity between the steady-states of the noisy and noiseless evolutions i.e., establishing that they belong to the same mixed-state phase~\cite{sang2023mixed}.

An important caveat is in order: although we characterize phases in terms of their steady-state order (or lack thereof), in any finite-sized system, there is always a non-vanishing probability that the system ends up in an absorbing configuration (i.e., with all erasure flags occupied). Once the dynamics lands in this configuration, the system cannot escape it; hence, in any finite-sized system, strictly speaking the $t\to\infty$ steady-state is always the absorbing state. However, the probability of reaching this fixed-point state can be exponentially small in the system size $L$~\cite{henkelNonEquilibriumPhaseTransitions2008}, such that the time-scale $\tau$ at which the absorbing state is reached increases exponentially with $L$. When we refer to a steady-state in this regime, we will thus always mean the steady-state of the system up to these exponentially long time-scales (which are beyond the range of our numerical simulations for all system sizes considered). Correspondingly, by a non-trivially ordered \textit{steady-state phase} we mean one that does not reach the absorbing state for times that are exponentially long in the system size. On the other hand, there exist regimes in parameter space where the absorbing state is reached on a time-scale (for large enough $L$) that is independent of system size. We refer to such regimes, for which the absorbing state is the only possible choice of steady-state, as being in the \textit{absorbing phase}.

\subsection*{Main results}

The central concept of our protocol is that it leverages information about the locations of stabilizer defects, provided by the heralded noise channel and encoded in the classical flags, to effectively confine the stabilizer defects using fully local dynamics. As a result of the competition between our local correction protocol and the heralded noise, the classical flags undergo an absorbing state transition from an active phase to the absorbing phase. The flag densities ($n^f_X$ and $n^f_Z$) serve as an order parameter for such transitions~\cite{hinrichsenNonequilibriumCriticalPhenomena2000,henkelNonEquilibriumPhaseTransitions2008}. Moreover, we show that the small values of the flag densities in the active phase precisely correspond to topological order in the steady state, as directly evidenced by low logical error rates in this phase.

To illustrate our protocol in a simple setting, we first consider the ground-space of the 2d Toric code subject to both local correction dynamics and unbiased heralded errors; the resulting steady-state phase diagram is shown in Fig.~\ref{fig:2dphase}(a). Our local correction protocol succeeds in stabilizing a topologically non-trivial mixed steady-state (the active phase in Fig.~\ref{fig:2dphase}) for noise below a certain threshold. Specifically, we show that this mixed-state (a) preserves logical information by applying the minimum weight perfect matching (MWPM) decoder, and (b) is connected to the original pure state by a local quantum channel whose depth scales at most logarithmically with the number of qubits. Thus, the active phase of our dynamics belongs to a mixed-state phase with $\mbZ_2$ TO. We also find that a first-order phase transition separates this active phase from neighboring dynamical phases, of which two (the $X$-active and $Z$-active phases in Fig.~\ref{fig:2dphase}(a)) preserve only classical information\footnote{These phases are characterized by the $\mbZ_2^{(0)}$ anyon theory and, in the parlance of Ref.~\cite{sohal2024imto}, constitute an intrinsically mixed TO.}. 

We then turn to our main focus, which is to show that local correction dynamics succeeds in stabilizing a mixed steady-state phase with $D_4$ TO for perfectly heralded noise (below a certain threshold). 
$D_4$ TO belongs to the general class of TOs which harbor point-like {\it non-Abelian} anyons: whereas Abelian anyons only acquire a universal phase when braided around each other, the internal degeneracy of non-Abelian anyons can be used to construct a richer set of quantum gates, which can be exploited for topological quantum computation~\cite{KitaevToric,nayakreview,cui2015,cong2017}. While states with arbitrary non-Abelian TO are challenging to prepare, those derived from solvable groups\footnote{Given a finite group $G$, its derived series is a set of normal subgroups $N_j$, derived inductively via $N_0 = G$ and $N_j = [N_{j-1},N_{j-1}]$ ($j>0$). Here, $[N,N]$ denotes the commutator subgroup of $N$. The derived length of $G$ is defined as the smallest positive integer $\ell_G$ such that $N_{\ell_G} = \mbZ_1$. A solvable group is defined as one for which $\ell_G$ is finite.} permit efficient preparation via adaptive quantum circuits~\cite{lu2022,nathierarchy,bravyi2022,tantivasadakarn2023shortest} and have been realized on NISQ devices~\cite{andersen2023,xu2023,iqbal2024nonabelian,fib1,goel2024}. In the specific case of $D_4$ TO, non-Abelian braiding was recently demonstrated on a trapped-ion processor~\cite{iqbal2024nonabelian}. Motivated in part by these experiments, we focus on TO in $d=2$ spatial dimensions in this paper. While we expect in principle that our protocol extends to $d=3$ spatial dimensions mutatis mutandis (since it is the point-like anyons that destabilize the phase, just as in $d=2$), we do not pursue this here since the numerical simulations are far more challenging.

Fig.~\ref{fig:2dphase}(b) shows the resulting phase diagram, which contains a fully active phase, an absorbing phase, and a partially active phase that preserves only a classical memory~\cite{sohal2024imto,ellison2024,zhang2024}. As for the Toric code, these are separated via first-order transitions. We provide strong evidence that the active phase of our local autonomous dynamics is in a mixed-state phase with $D_4$ TO: we show both that the MWPM decoder faithfully recovers quantum information and verify numerically that the steady-state is connected to a pure $D_4$ ground state via a local quantum channel whose depth scales at most logarithmically with the number of system qubits. We will also show that while the steady-state TO is unstable against imperfectly heralded errors, when most errors are heralded our protocol leads to a significant enhancement in the lifetime of the encoded logical information. 

At first blush, it is rather surprising that local error correction succeeds for a non-Abelian anyon theory: the string-operators that are required to pair-annihilate non-Abelian anyons are unitary operators whose depth is lower bounded by the linear separation between the two anyons~\cite{shi2019}. Na\"ively, this suggests that local error correction must fail. Our approach circumvents this by exploiting the fact that in anyon theories given by the quantum double $\mathcal{D}(G)$~\cite{dijkgraaf,KitaevToric}, where $G$ is a class-2 nilpotent group (e.g. $D_4$), non-Abelian anyons of the same type fuse to Abelian anyons. Due to this nilpotency (or acyclicity~\cite{galindo2018}), non-Abelian anyons can be moved via depth-$1$ unitary circuits at the expense of leaving behind a trail of Abelian anyons, which are pair-created at the end points of Pauli strings. In correcting for potential non-Abelian errors, our correction protocol allows partial information about where such strings are created, thereby enabling the confinement of both Abelian and non-Abelian errors in the long-time steady-state at noise rates below a certain threshold.

A technical contribution of this work is a generalization of the stabilizer tableau~\cite{gottesman1998heisenberg,aaronson2004improved} to the case of ``quasi-stabilizers" (defined below), which permits efficient simulation of Lindblad dynamics for a certain class of non-Pauli stabilizers. 

Finally, we comment on the values of correction rates required for these protocols and their feasibility on current erasure qubit platforms. We imagine a setup where each qubit gets acted on by $m$ gates per unit time for purposes of some desired quantum information processing task. We assume that each of these gates fails with probability $p_g$, resulting in a heralded error on the qubit. For small values of $p_g$, this can be modeled as an error process with rate $\eta_g = m p_g$. At the same time, the qubit is also affected by entangling gates that are applied during the measurement-feedback routine of our local correction protocol. The value of $\gamma$ shown in Fig.~\ref{fig:2dphase} indicates the rate at which the protocol must check for flags; however, the measurement and feedback gates are performed only when the classical erasure flags are in an appropriate correctable configuration. This rate is given by $\alpha \gamma$ ($\alpha<1$), with $\alpha\approx0.03$ in the active phase of the $D_4$ model (see Fig.~\ref{fig:d4-a-checks}). If the measurement circuit fails with probability $p_c$, then the measurements during the local correction routine themselves contribute to the noise, with a rate $\eta_c = \alpha \gamma p_c$\footnote{Here, we are implicitly assuming that errors in the measurement-feedback routine have a qualitatively similar effect to applying a single-site Pauli error of the type analyzed in our model immediately following the measurement-feedback circuit.}. Finally, qubits remaining in the idling state can decay from the computational subspace, leading to idling noise rate $\eta_{\text{idle}}$. Combining these contributions, we expect the total noise rate to be 
given by: \eq{eq:total-noise-num}{\eta=\eta_{\text{idle}}+m p_g + \alpha \gamma p_c.} 
 
As shown in Fig.~\ref{fig:2dphase}(b), $D_4$ topological order can be stabilized provided that $\gamma/\eta = R\approx 200$. Since the rates must be non-negative, Eq.~\eqref{eq:total-noise-num} can be satisfied only if $p_c <\frac{1}{R \alpha }$. For Yb-171 atoms~\cite{maHighfidelityGatesMidcircuit2023}, erasure qubits were recently demonstrated with $p_g\sim 0.02$ for two-qubit gates; taking $R\alpha \approx 10$ allows values of $p_c$ up to $0.1$, meaning that measurement-feedback must be achieved with less than 5 two-qubit gates in order to access the correction regime. In these systems $\eta_{\text{idle}}\sim 0.3 s^{-1}$, while gate times are on the order of $\mu s$, suggesting that, provided the measurement error is sufficiently small, our protocol can easily control the idling errors.


\section{Warm-up: Self-correcting 2d Toric code with heralded noise}
\label{sec:TC}

Before discussing our correction protocol for the non-Abelian $D_4$ model, we begin by illustrating how such a protocol works for the relatively simpler case of the celebrated 2d Toric code~\cite{KitaevToric}, which hosts (Abelian) $\mbZ_2$ topological order and has been experimentally realized on various NISQ platforms~\cite{satzinger2021,semeghini2021,fossfeig2023,iqbal2023}.

\begin{figure}
\includegraphics{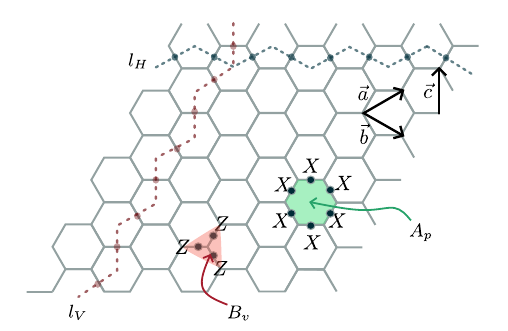}
\caption{\label{fig:tc-1}2d Toric code on the honeycomb lattice: qubits live on the edges of a honeycomb lattice. A patch of $12\times 12$ size lattice is shown, with PBC in vertical and horizontal directions implemented according to Eq.~\eqref{eq:tc-pbc}. Edges involved in the vertex stabilizer $B_v$ and plaquette stabilizer $A_p$ are shown in red and green, respectively. The operators $l_H$ and $l_V$ run along non-contractible loops on the dual triangular lattice, defining the logical $\mathcal{Z}_H$ and $\mathcal{Z}_V$ operators, respectively. } 
\end{figure}

We consider a honeycomb lattice defined on a two-dimensional torus, with a qubit (together with an $X$ and a $Z$ flag) placed on each edge.  The periodic boundary conditions (PBC) on the $L_x\times L_y$ sized torus are implemented by making the following identifications for the position vector $\vec{r}$: 
\eq{eq:tc-pbc}{\vec{r} \sim \vec{r}+\frac{n_x}{3}L_x (\vec{a}+\vec{b})+\frac{n_y}{3}L_y (\vec{a}+\vec{c})}
for $n_x,n_y \in \mbZ$ and where $\vec{a}, \vec{b}$, and $\vec{c}$ are the basis vectors of the honeycomb lattice, as shown in Fig.~\ref{fig:tc-1}; these are related by $\vec{a}=\vec{b}+\vec{c}$. In the following, we set $L_x=L_y=L$, which results in a  lattice with $L^2$ edges, $L^2/3$ plaquettes, and $2L^2/3$ vertices. Note that the Toric code can be defined on any 2D lattice; we choose the honeycomb lattice here to make contact with the $D_4$ model discussed later. 

The Toric code ground states $\ket{\psi}$ are those states that are simultaneously stabilized by a set of vertex $B_v$ and plaquette $A_p$ stabilizers i.e., they satisfy $B_v \ket{\psi} = A_p \ket{\psi} = \ket{\psi} \, \forall \, v,p$, with
\eq{Eq:TCStabs}{ B_v = \prod_{j\in ^*v} Z_j \, ,\qquad A_p = \prod_{k\in \partial p} X_k \, .}
Here $^*v$ contains the three edges attached to vertex $v$, and $\partial p$ consists of the six edges surrounding plaquette $p$. On the 2-torus, there are four locally indistinguishable ground states which can be characterized by the eigenvalues of the closed Wilson loop operators $\mathcal{Z}_H \equiv \prod_{j\in l_H}Z_{j} = \pm 1, \mathcal{Z}_V \equiv \prod_{j\in l_V} Z_{j}=\pm1$, where $l_H (l_V)$ is a horizontal (vertical) non-contractible loop on the dual lattice (see Fig.~\ref{fig:tc-1}).

\subsection{Noise model for unbiased heralded errors}
\label{sec:tcnoise}

We first introduce the noise model that we will use throughout this work, which consists of unbiased, heralded errors. To model a system with heralded errors, in addition to the qubits themselves, we also introduce a classical degree of freedom, which we refer to as an erasure ``flag". These flags are represented by a pair of occupation numbers $n^f_{X_j}$ and $n^f_{Z_j}$ which are associated with each qubit $j$ and can independently be in either occupied ($n^f_{X}=1,n^f_{Z}=1$) or unoccupied ($n^f_{X}=0,n^f_{Z}=0$) state. Here, we will set both $X$ and $Z$ flags on all qubits to be unoccupied $n^f_{X_j} =n^f_{Z_j} =0 \, \forall \, j$ in the initial state.  

Here we will use an error model that describes unbiased heralded noise, for which (after Pauli twirling~\cite{wallman2016pauliTwirl}) an erasure at site $j$ acts on the qubits via the local quantum channel:
\eq{Eq:DepolNoise}{ \rho\rightarrow \mathcal{E}_j(\rho) = \frac{1}{4}\rho+ \frac{1}{4}X_j\rho X_j + \frac{1}{4}Y_j\rho Y_j +\frac{1}{4} Z_j\rho Z_j \, . }
While erasures typically lead to a more restricted set of errors in experiments, here we will show that 2d TOs can be dynamically stabilized without requiring any bias in the noise channel --  this is in contrast with 1d SPTs, where the strong symmetry constraint necessitated biased erasure noise~\cite{chirame2024spt}. Indeed, as we will shortly demonstrate, for the 2d Toric code our protocol works comparably well for biased and unbiased noise. For the $D_4$ case studied later, we will see that unbiased heralded noise represents a worst-case noise scenario for our decoder; we hence expect that more realistic noise models will lead to a significantly lower decoding overhead. 

Because the errors are heralded, our noise channel also acts on the flags: upon detection of an erasure at qubit $j$, both $X$ and $Z$ flags at this site are raised by setting $n^f_{X_j} =1$ and $n^f_{Z_j} =1$. We model the action of this noise channel using continuous time Lindbladian evolution~\cite{lindblad1976generators, gorini}, given by 
\eq{Eq:LindbladNoise}{ \mathcal{L}_{\eta}[\rho] = \sum_{\alpha,\mathsf{q_x},\mathsf{q_z},j} L_{\alpha,j}^{(\mathsf{q_x},\mathsf{q_z})} \rho L^{\dag(\mathsf{q_x},\mathsf{q_z})}_{\alpha,j}- \frac{1}{2} \{ L^{\dag(\mathsf{q_x},\mathsf{q_z})}_{\alpha,j} L_{\alpha,j}^{(\mathsf{q_x},\mathsf{q_z})}, \rho \} \, ,
}
where the Lindblad jump operators associated with the noise channel are
\eq{eq:tc-noise}{
L_{\alpha,j}^{(\mathsf{q_x},\mathsf{q_z})} =  \sqrt{\frac{\eta}{4}}  f^+_{X_j} (f^-_{X_j})^\mathsf{q_x} f^+_{Z_j} (f^-_{Z_j})^ \mathsf{q_z} \ \hat{\sigma}^\alpha_j 
\, .}
Here, $j$ labels the qubit index and $\mathsf{q_x,q_z}\in\{0,1\}$. The  operators  $f^+_{X_j} (f^-_{X_j})$ raise $n^f_{X_j}$ from $0$ to $1$ (lower  $n^f_{X_j}$ from $1$ to $0$). The relation $f^+_{X_j} f^-_{X_j} = n^f_{X_j}$ implies that the jump operator labeled by $\mathsf{q_x}=1 (\mathsf{q_x}=0)$ acts on the site if an $X$ flag was already present (absent). The operators $f^{\pm}_{Z_j}$ act analogously on the $Z$ flags. Here
\eq{}{ \hat{\sigma}^0_j \equiv \mathbf{1} \ , \ \ \hat{\sigma}^1_j \equiv X_j \ , \ \ \hat{\sigma}^2_j \equiv Y_j \ , \ \ \hat{\sigma}^3_j \equiv Z_j  
}
are standard Pauli operators acting on the qubit at site $j$. The $\alpha = 0$ terms in the noise channel represent processes in which the qubit is miraculously initialized in the correct state; in this case the only effect of the erasure is to raise any flags that were not already raised at the relevant site. The remaining terms represent processes in which both flags are raised (if not already present) and a Pauli error occurs.


\subsection{Correction dynamics for the Toric code}
\label{sec:tccorrect}

A key feature of the Toric code stabilizers in Eq.~\eqref{Eq:TCStabs} is that vertex (plaquette) stabilizer defects are necessarily created in pairs, and are spatially separated via a string of contiguous Pauli-$X$ (Pauli-$Z$) errors. This is a robust feature of the model because the stabilizer defects correspond to emergent anyons, which cannot be locally created.  For perfectly heralded errors, this means that pairs of stabilizer defects are necessarily connected by strings of flags. Our correction protocol uses the information provided by the positions of these flags to appropriately move stabilizer defects along such strings, thereby annihilating each stabilizer defect with its partner. 
\begin{figure}
\includegraphics{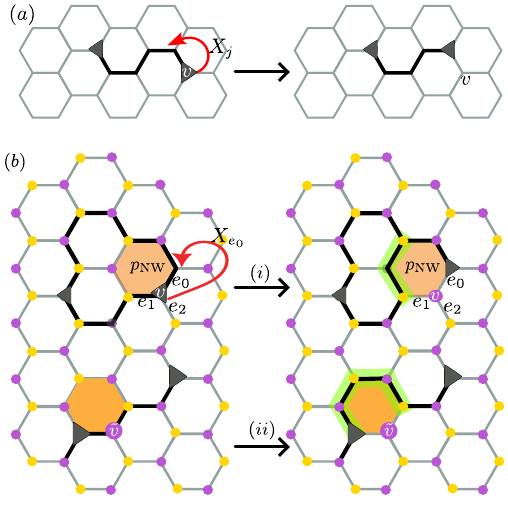}
\caption{\label{fig:tc-x-crc} $X$ correction protocol for Toric code: The edges that have $X$ flags on them are shown using bold lines, and the vertex defects are depicted by black triangles. The initial configurations on the left side are updated to obtain the configuration on the right side. We illustrate processes in which stabilizer defects are present near the corrected vertex; the processes with no stabilizer defects act identically on the flags, but do not act on the qubits.  (a) X-leaf move: Vertex $v$ has a single edge with an $X$ flag on it, and hosts a stabilizer defect. The defect is displaced by applying $X_j$ on the flagged edge $j$, after which the $X$ flag is removed. (b) X-loop move: (i) The two edges ($e_0$ and $e_1$) emanating from a vertex $v$ of the purple sublattice and bordering its northwest plaquette $p_{NW}$ have $X$ flags on them. The correction step in Eq.~\eqref{eq:tc-x-loop} removes these flags, raises new $X$ flags on the remaining edges of $p_{NW}$ (if they were not present to begin with; highlighted in green color), and displaces the vertex defect by applying $X_{e_0}$. (ii) The X-loop move can also be applied at vertex $\widetilde{v}$ which locally hosts a flag configuration $n^f_{X_{e_0}}=n^f_{X_{e_1}}=1$, and $n^f_{X_{e_2}}=0$ even though the flags do not form a closed loop. As a result of this operation, the vertex defects will be paired along the alternative path highlighted in green color. The loop moves are applied only at the vertices on the purple sublattice.} 
\end{figure}
\paragraph{$X$ flags and $B_v$ defects:} To see how the correction protocol works in practice, let us consider a vertex stabilizer defect $B_v=-1$. If only one of the three edges (say edge $j$) surrounding this defect has an $X$ flag, then we can be certain that the defect is the result of an $X$ type error that occurred at this particular qubit. Consequently, we move the defect by applying an $X$ operator to this qubit and simultaneously lower the $X$ flag, as illustrated in Fig.~\ref{fig:tc-x-crc}(a). Similarly, at a vertex with no stabilizer defect and where only one of the edges has an $X$ flag, we can simply erase the flag. These moves, which we call {\it leaf moves}, are implemented via the local jump operators:
\eq{eq:tc-x-leaf}{L^{\text{TC}^{\text{X-leaf}}}_{v,j,s} = \frac{\sqrt{\gamma_x}}{2}f^-_{X_j} X^{\frac{1-s}{2}}_{j} (1+s B_v) \prod_{k\in ^*v\setminus j} (1-n^f_{X_k}) \, .}
Here, $v$ labels the $2L^2/3$ vertices of the honeycomb lattice and $k\in ^*v\setminus j$ denotes the two edges $\neq j$ attached to vertex $v$. The scalar $s=-1 (s=+1)$ corresponds to the case where a vertex defect is present (absent), and $\gamma_x$ parametrizes the local correction rate. The dynamics generated by this jump operator autonomously implements measurement and feedback towards a defect-free state (and requires no postselection).

In one spatial dimension, where there can be no contractible closed loops of $X$ flags, these two processes are sufficient for stabilizing a topologically non-trivial steady-state, as we previously demonstrated in Ref.~\cite{chirame2024spt}. However, because the leaf moves cannot remove closed loops of $X$-type flags, in two spatial dimensions we require additional moves to stabilize a topologically ordered steady-state. To this end, we introduce the $X$-\textit{loop move}, shown in Fig.~\ref{fig:tc-x-crc}(b), which moves the stabilizer defects along a loop and erases the corresponding loop segment in a process which is reminiscent of Toom's rule~\cite{toom1980stable}.

Consider the configuration of flags and defects shown in the left panel of Fig.~\ref{fig:tc-x-crc}(b). 
Here, the vertex $v$ on the purple sublattice is found with flags on the two edges $e_0$ and $e_1$ that are North or West of it, and no flag on the third edge $e_2$.  Our protocol detects such configurations by checking if the flag configuration is of the form $n^f_{X_{e_0}}=n^f_{X_{e_1}}=1$ and $n^f_{X_{e_2}}=0$. When such a configuration is detected, the protocol lowers the flags on $e_0$ and $e_1$, and raises new $X$ flags on the outer edges of the North-Western plaquette $p_{NW}$ (if they are not already present, as is the case for the vertex $\widetilde{v}$ in the Figure). The $B_v$ stabilizer at the central vertex is then measured, and if $B_v=-1$, the defect is moved up and to the right by applying $X_{e_0}$. The overall result of this operation is to push the string or loop of $n^f_X=1$ flags in the North-West direction. Concurrently, the vertex defects are (a) either paired-up along the path or (b) the newly raised flags provide an alternate path on the Honeycomb lattice that is equivalent to the original error-string up to a product of plaquette stabilizers $A_p$. Since the purpose of these moves is to shrink the loop in a preferential direction, the $X$-loop moves are not applied to vertices on the yellow sublattice. At a vertex $v$ on the purple sublattice, this operation is implemented by the local jump operators:
\eq{eq:tc-x-loop}{\eqsp{L^{\text{TC}^{\text{X-loop}}}_{v,s,\mathbf{m^x}}=\frac{\sqrt{\gamma_x}}{2} &  X^{(1-s)/2}_{e_0}\prod_{j\in(\partial p_{NW}\setminus ^*v)}f^+_{X_j}(f^-_{X_j})^{\mathbf{m^x}_j}\\
& (1+s B_v) f^-_{X_{e_0}} f^-_{X_{e_1}}(1-n^f_{X_{e_2}}) \, .}}
Here, $\mathbf{m^x}=\{0,1\}^{\otimes 4}$ is a length-4 bit string where each entry corresponds to the locations of the four qubits where potentially new $X$ flags are raised. As a result of relation $f^+_Xf^-_X=n^f_X$, the jump operators for which $\mathbf{m^x}_j=1(\mathbf{m^x}_j=0)$ act non-trivially when an $X$ flag is already present (absent). The index $s=\pm1$ corresponds to the measurement outcome of $B_v$. As before, $\partial p$ and $^*v$ represent the qubits on the edges bordering the plaquette $p$ and connected to the vertex $v$, respectively.

\begin{figure*}
\includegraphics{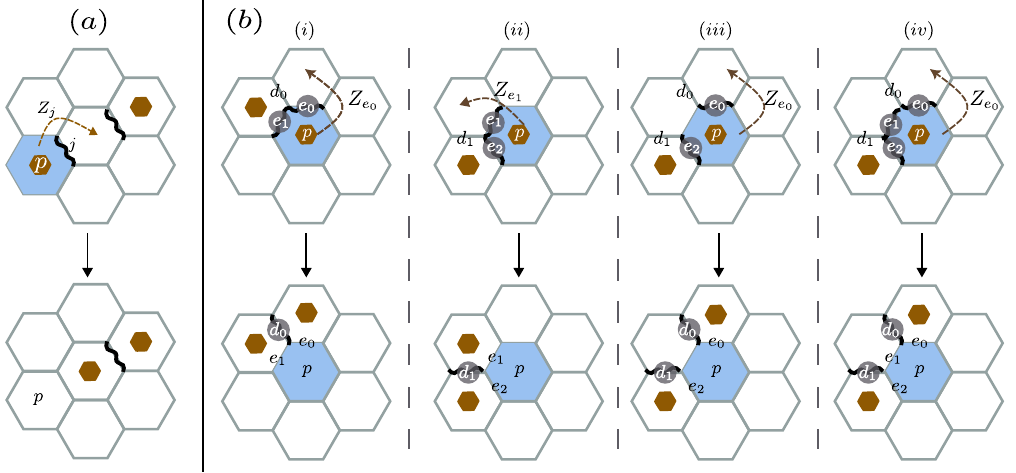}
\caption{\label{fig:tc-z-crc} $Z$ correction protocol for Toric code: The edges that have $Z$ flags are shown using bold wavy lines. The correction operation is applied to the blue plaquette in the top row to obtain the new configuration in the bottom row. Here, we illustrate processes in which stabilizer defects are moved by acting with $Z$ on appropriate edges (shown using dashed arrows); the processes with no stabilizer defects act identically on the flags, but do not apply the Pauli-$Z$ operator. (a) Z-leaf move: a plaquette defect occurs on a plaquette with a single flagged edge. The correction protocol applies $Z$ to this flagged edge, and subsequently erases the flag, moving the defect to a neighboring plaquette. (b) Z-loop moves: Four candidate processes corresponding to $Z$ flags on edges labeled by $(e_0,e_1)$,$(e_1,e_2)$,$(e_0,e_2)$, and $(e_0,e_1,e_2)$ are shown here. In each configuration, new $Z$ flags are added on edges $d_0$ and/or $d_1$ as described in the main text (refer to Eq.~\eqref{eq:tc-z-loop}).} 
\end{figure*}
\paragraph{$Z$ flags and $A_p$ defects:} A similar process on the dual triangular lattice, coupled to $Z$ flags, can be used to pair-annihilate plaquette stabilizer defects. First, if only one out of six edges surrounding the plaquette $p$ has a $Z$ flag (refer to Fig.~\ref{fig:tc-z-crc}(a)), a $Z$ leaf move is performed using the Lindblad jump operators:
\eq{eq:tc-z-leaf}{L^{\text{TC}^{\text{Z-leaf}}}_{p,j,s} = \frac{\sqrt{\gamma_z}}{2}f^-_{Z_j} Z^{\frac{1-s}{2}}_j (1+s A_p) \prod_{k\in \partial p\setminus j}(1- n^f_{Z_k}) \, ,
}
where $\gamma_z$ sets the rate of $Z$ correction. This leaf move effectively moves a plaquette defect towards its partner while lowering $Z$ flags along the error string connecting the two defects.

As with $X$ flags, loop moves are also necessary to remove $Z$ flags that form closed loops on the dual lattice. Let us consider a plaquette $p$ that has more than one $Z$ flag on its North-Western edges. The possible configurations can be labeled by the occupied edges $\mathcal{E}=\{ (e_0,e_1),(e_1,e_2),(e_0,e_2),(e_0,e_1,e_2)\}$, as illustrated in Fig.~\ref{fig:tc-z-crc}(b). For each configuration, the figure describes the locations $R[\mathcal{E}]=\{(d_0),(d_1),(d_0,d_1),(d_0,d_1)\}$ of new $Z$ flags which are consistent with this network, meaning that errors on these edges are equivalent to errors on the original flagged edges, up to a product of vertex stabilizers $B_v$. For a given $\varepsilon\in \mathcal{E}$ and plaquette $p$, the corresponding jump operator is:
\eq{eq:tc-z-loop}{\eqsp{
L^{\text{TC}^{\text{Z-loop}}}_{p,s,\varepsilon,\mathbf{m^z}} =& \frac{\sqrt{\gamma_z}}{2} Z^{(1-s)/2}_{\varepsilon^{(0)}} \prod_{k\in R[\varepsilon]} f^+_{Z_k}(f^-_{Z_k})^{\mathbf{m^z}_k}  (1+sA_p)\\
& \prod_{j\in\varepsilon}f^-_{Z_j}\prod_{l\in(\partial p\setminus\varepsilon)}(1-n^f_{Z_l}) \, .}}
Here, $s=\pm1$ denotes the stabilizer measurement outcome and $\mathbf{m^z}\in\{0,1\}^{\otimes \text{dim}(R[\varepsilon])}$ is bit-string that accounts for all possible configurations of $Z$ flags on the edges in $R[\varepsilon]$. Upon detection of a plaquette defect, i.e. $s=-1$, it is always moved towards the north-most neighbor which shares the edge $\varepsilon^{(0)}$ with $p$.

The $X$ ($Z$) loop moves allow the dynamics to shrink contractible loops of flags that may span multiple plaquettes (or vertices, for $Z$ flags).  By choosing a particular direction in which to ``push" loop segments, this can be accomplished using the sequence of local moves acting on a single plaquette (vertex) and the edges in its immediate vicinity. Evidently, this also means that the loop moves do not act {\it only} on loops; in certain configurations, they will instead re-arrange the local flag and stabilizer defect configuration, as shown in Fig.~\ref{fig:tc-x-crc}(b-ii). Note, however, that each of these moves only involves an $O(1)$ number of flags and stabilizer operators, thereby retaining the local nature of the dynamics.

The total superoperator implementing the correction processes is then given by
\eq{eq:LindbladTCCorr}{\begin{aligned}\mathcal{L}_C &=
\ \gamma_x \sum_{\substack{v,s=\pm1\\j\in^*v}} \mathcal{L}^{\text{TC}^{\text{X-leaf}}}_{v,j,s}     \ + \ \gamma_z \sum_{\substack{p,s=\pm1\\ j\in\partial p}} \mathcal{L}^{\text{TC}^{\text{Z-leaf}}}_{p,j,s}     \\
& + \gamma_x \sum_{\substack{v\in{\text{purple}},s=\pm1\\\mathbf{m^x}\in\{0,1\}^{\otimes4}}} \mathcal{L}^{\text{TC}^{\text{X-loop}}}_{v,s,\mathbf{m^x}}     
+\gamma_z\sum_{\substack{p,s=\pm1,\\ \varepsilon\in\mathcal{E}(p),\mathbf{m^z}}}\mathcal{L}^{\text{TC}^{\text{Z-loop}}}_{p,s,\varepsilon,\mathbf{m^z}}  ,
\end{aligned}}
where the individual superoperators have the standard Lindblad form $\gamma_{\Phi}\mathcal{L}_{\Phi}[\rho]=L_{\Phi}\rho L_{\Phi}^\dag-\frac{1}{2}\{L^\dag_{\Phi} L_{\Phi},\rho \}$ and the jump operators are defined in Eqs.~\eqref{eq:tc-x-leaf}-\eqref{eq:tc-z-loop}. The complete dynamics is given by time evolving the density matrix under both the erasure noise Eq.~\eqref{Eq:LindbladNoise} and the correction dynamics Eq.~\eqref{eq:LindbladTCCorr}, in accordance with the Lindblad master equation:
\eq{eq:total-lindblad}{\frac{d}{dt}\rho(t) = \mathcal{L}_\eta[\rho(t)] + \mathcal{L}_C[\rho(t)] \, .}

\begin{figure*}
\includegraphics[width=\textwidth]{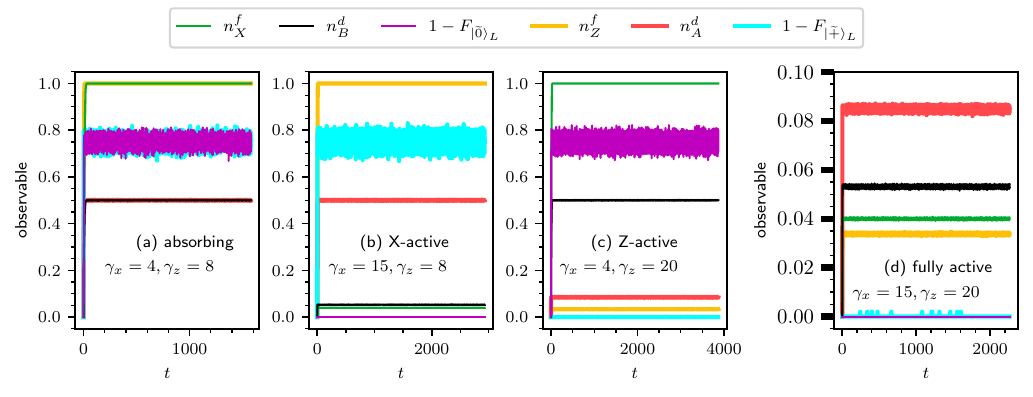}
\caption{\label{fig:TC-MC-dynamics} 
Time evolution of flag and defect densities, together with the logical fidelity, for the Toric code on a $48 \times 48$ honeycomb lattice for four representative points in the $\gamma_x$ - $\gamma_z$ phase diagram (see Fig.~\ref{fig:2dphase}(a)). The probabilities of logical $X$ errors ($\equiv 1-F_{|\widetilde{0}\rangle_L}$) and logical $Z$ errors ($\equiv 1-F_{|\widetilde{+}\rangle_L}$) are computed by decoding the state at time $t$ using MWPM decoder (see Eqs.~\eqref{eq:tc-logical-states},\eqref{eq:tc-fidelity}). The noise rate is set to $\eta=1$, with each data point representing an average over $500$ Monte Carlo runs. (a) Absorbing phase $(\gamma_x=4,\gamma_z=8)$: where flags rapidly reach their absorbing state $n^f_X=n^f_Z=1$, and all qubits become maximally mixed. The logical error rates $1-F_{|\widetilde{0}\rangle_L}$ and $1-F_{|\widetilde{+}\rangle_L}$ both saturate to their maximal value. (b) Partially X-active phase $(\gamma_x=15,\gamma_z=8)$: The density of $X$ flags and $B$ defects saturates at a small value, whereas $Z$ flags and $A$ defects proliferate, leading to small (large) values of logical $X (Z)$ error rate. (c) The analogous partially Z-active phase $(\gamma_x=4,\gamma_z=20)$ (d) Active phase $(\gamma_x=15,\gamma_z=20)$: here, both $X$ and $Z$ flag densities saturate to small values. Both logical $X$ (magenta colored curve) and logical $Z$ (cyan colored curve) error rates remain close to zero, and hence the quantum information can be recovered up to times exponentially large in system size.}
\end{figure*}


\subsection{Metrics for topological order and numerical results}
\label{sec:tcmetrics}

Our goal is to use the local correction dynamics to stabilize a steady-state that belongs to the same mixed-state phase as the Toric code ground state. Practically, this means that the error strings must remain short, such that the stabilizer defects are effectively confined. One consequence of such confinement is that a minimum-weight perfect matching (MWPM) error correction decoder~\cite{dennis2002} will not produce logical errors (assuming that the confinement scale is small compared to the system size). In the following, we will use this decodability as an effective order parameter for the topological phase.

An alternative metric for our steady-state is provided by the recent equivalence relation placed on the space of density matrices in Ref.~\cite{sang2023mixed}. Namely, two density matrices $\rho_1$ and $\rho_2$ belong to the same mixed-state phase if they are two-way connected via finite depth quasi-local quantum channels i.e., if there exist quasi-local channels $\Sigma_{12}$ and $\Sigma_{21}$ such that $\Sigma_{21}(\rho_1) = \rho_2$ and $\Sigma_{12}(\rho_2) = \rho_1$, where the depth of these channels is at most $\mathcal{O}(\text{polylog}(L))$. In the particular case of the 2d Toric code, it is known that there exists a local decoder whose threshold is quantitatively extremely close to that of the MWPM decoder~\cite{sang2023mixed}; thus, here we focus only on the MWPM decoder and defer further analysis of this second metric to the $D_4$ active phase considered in Sec.~\ref{sec:phases}.

The full Hilbert space of our model,  consisting of both qubits and flags, is spanned by the combined eigenstates of the $A_p,B_v$ stabilizers and the flag occupation-numbers. The noise and correction Lindbladians in Eq.~\eqref{eq:total-lindblad}, which generate the total dynamics, do not couple the diagonal and off-diagonal elements of the density matrix $\rho$ in this basis. Hence, the time evolution can be simulated as a stochastic Monte Carlo dynamics of stabilizer and flag populations~\cite{hinrichsenNonequilibriumCriticalPhenomena2000}. We initialize the system in the Toric code ground state subspace, where $B_v=A_p=+1$ for all vertices $v$ and plaquettes $p$, with the flags set to $n^f_{X_j}=n^f_{Z_j}=0$ on all edges $j$. We then simulate the dynamics using random sequential updates of the state, where the rates of all processes are specified in the previous section. The time-evolved observables are computed by averaging over multiple independent Monte Carlo runs (see Appendix.~\ref{app:numerics} for additional details). In what follows, we set the noise rate of the erasure channel to $\eta=1$ and explore the steady-state phase diagram as a function of the $X$ and $Z$ correction rates $\gamma_x$ and $\gamma_z$, respectively.

Fig.~\ref{fig:2dphase}(a) shows the numerically obtained steady-state flag density as a function of the two correction rates. As is clear from the figure, tuning the correction rates leads to four different phases: a fully \textit{active phase} for large values of $\gamma_x$ and $\gamma_z$, where the density of both types of flags saturates to a small value; an \textit{absorbing phase} when both correction rates are small, where both $X$ and $Z$ flags percolate and saturate to steady-state values of $n^f_X=n^f_Z=1$; and two \emph{partially active phases}, where only one of the two types of flags percolates, while the other saturates to a low-density in the steady-state.

In the active phase, the stabilizer defects are corrected along the strings of erasures frequently enough such that the density of defects remains small, as shown in Fig.~\ref{fig:TC-MC-dynamics}(d) at a representative point in the phase. Here, the densities of the vertex and plaquette stabilizer defects are defined, respectively, as
\be
n_B^d = \frac{3}{2L^2} \sum_v \frac{\left(1 - \braket{B_v} \right)}{2} \, , \,\, n_A^d = \frac{3}{L^2} \sum_p \frac{\left(1 - \braket{A_p} \right)}{2} \, .
\ee
The heralded nature of the noise and the correction dynamics also ensures that pairs of vertex (plaquette) defects are necessarily connected by strings of $X$ ($Z$) type flags. Hence, a low density of flags necessarily corresponds to confinement of the stabilizer defects.  We further illustrate this by evaluating the probability of a logical error upon pairing the defects in the time-evolved state using the MWPM decoder, ignoring the configuration of the flags. The logical state (i.e., the exact initial state within the ground state Hilbert space) is recovered by this decoder if the probability of both logical $X$ errors (logical bit-flips) and logical $Z$ errors (logical phase-flips) remains low. These error probabilities can be estimated by initializing the system in two different basis states in the Toric code ground-state space, defined as 
\eq{eq:tc-logical-states}{\eqsp{\mathcal{Z}_V|\widetilde{0}\rangle_L&=\mathcal{Z}_H|\widetilde{0}\rangle_L=|\widetilde{0}\rangle_L\\
\mathcal{X}_V|\widetilde{+}\rangle_L&=\mathcal{X}_H|\widetilde{+}\rangle_L=|\widetilde{+}\rangle_L
}}
where $\mathcal{Z}_{V/H}$ are the Wilson line operators shown in Fig.~\ref{fig:tc-1} (which also serve as logical Pauli-$Z$ operators), and the logical Pauli-$X$ operators $\mathcal{X}_{V/H}$ are equivalent line operators comprised of products of $X$ operators along closed vertical and horizontal curves on the direct lattice.    
The state of the qubits in a particular Monte Carlo trajectory $r$ initialized in $|\psi_{t=0}\rangle$ can be written as
\eq{}{|\psi_r\rangle = \prod_{j\in e^{(r)}_x} X_j\prod_{k\in e^{(r)}_z}Z_k|\psi_{t=0}\rangle \, .}
Here, $e^{(r)}_x(e^{(r)}_z)$ denotes the set of edges where Pauli $X$ (Pauli $Z$) noise is applied (up to this time) as a result of the Lindblad dynamics. The MWPM decoder generates the set of edges $d^{(r)}_x,d^{(r)}_z$ with smallest weight (counted in terms of number of non-identity Pauli operators) that pair-up all the defects such that the decoded state
\eq{}{|\psi^{(r)}_{\text{decoded}}\rangle := \prod_{j\in (e^{(r)}_x\oplus d^{(r)}_x)} X_j\prod_{k\in (e^{(r)}_z\oplus d^{(r)}_z)}Z_k|\psi_{t=0}\rangle}
has no stabilizer defects. If the edges in $e^{(r)}_x\oplus d^{(r)}_x$ ($e^{(r)}_z\oplus d^{(r)}_z$) form an odd number of non-contractible loops around the torus, they act as a logical $X$ (logical $Z$) operator on the state. Since $\{\mathcal{Z}_V,\mathcal{X}_H\}=\{\mathcal{Z}_H,\mathcal{X}_V\}=0$, such non-contractible loops flip the eigenvalue of the logical $Z$ (logical $X)$ operator relative to the initial state, thereby taking $|\widetilde{0}\rangle_L \leftrightarrow |\widetilde{1}\rangle_L$ ($|\widetilde{+}\rangle_L \leftrightarrow |\widetilde{-}\rangle_L$). We characterize the probability of such logical flips in terms of fidelity of decoded state with respect to the initial state as
\eq{eq:tc-fidelity}{F_{|\psi_{t=0}\rangle} = \frac{1}{N_r}\sum_{r=1}^{N_r}|\langle\psi_{t=0}|\psi^{(r)}_{\mathrm{decoded}}\rangle|^2,}
where $N_r$ is the number of independent Monte Carlo realizations. We numerically determine the state of logical information by pairing the stabilizer defects in the time-evolved state using the MWPM decoder implemented in the PyMatching-2 package~\cite{higgott2023pymatching}. The probabilities of logical $X$ errors ($\equiv 1-F_{|\widetilde{0}\rangle_L}$) and logical $Z$ errors ($\equiv 1-F_{|\widetilde{+}\rangle_L}$) are the computed in terms of the fidelity of this decoded state and two distinct initial states defined in Eq.~\eqref{eq:tc-logical-states}. In Fig.~\ref{fig:TC-MC-dynamics}(d), we see that the probability of both types of logical flips is essentially zero in the active state. The small values of logical $X$ ($Z$) error rates equivalently implies the protection of initially encoded logical $Z$ ($X$) information. This provides explicit evidence that quantum information is protected by the dynamical correction protocol for long times. 

In contrast, the absorbing state is the unique \emph{dead state} for the flags: when the flag density is $1$, none of our correction operations can act on any sites, and the dynamics consequently no longer acts on the flags. Once the flags enter the absorbing state, they can not leave this configuration via any of the dynamical processes. In this state, since all the correction jump operators vanish identically, the qubits evolve under the depolarizing channel and the dynamics leads to a maximally mixed steady state. This is evident in Fig.~\ref{fig:TC-MC-dynamics}(a), which shows that the stabilizer defect density in this phase saturates to $1/2$. All encoded information is consequently lost, as is evident from the fact that the probability of logical flips saturates to $1-\frac{1}{2^2}=0.75$ since the decoder returns all possible outcomes with equal probability. 

In the partially $X$-active ($Z$-active) phase, only $Z$ (only $X$) flags have proliferated. Thus, our correction protocol no longer corrects $Z$ ($X$) errors, but remains effective on $X$ ($Z$) errors. One manifestation of this is that only logical $Z$ (logical $X$) information is protected in the steady-state, as shown in Fig.~\ref{fig:TC-MC-dynamics}(b) and (c). Each of these phases thus represents a  {\it classical} memory, in which the information encoded in only one type of logical operator is dynamically protected. These phases represent intrinsically mixed topological phases which cannot be realized in the ground-states of 2d local gapped Hamiltonians~\cite{sohal2024imto,zhang2024} since they cannot be two-way channel connected to the trivial state on the 2-torus (even though they are two-way channel connected on the infinite plane).


\section{Dynamical error correction protocol for $D_4$ topological order}
\label{sec:D4}

We now turn to the main focus of this work -- namely, exploiting heralding to achieve a steady-state phase that harbors $D_4$ non-Abelian topological order (TO) in the presence of continuous noise processes. We focus on $D_4$ TO, whose anyon content is described by the quantum double $\mathcal{D}(D_4)$, for two reasons: first, this TO permits efficient preparation on a quantum processor using measurements and unitary feedback~\cite{nathierarchy,tantivasadakarn2023shortest,bravyi2022}, as has been demonstrated experimentally~\cite{iqbal2024nonabelian}. Second, despite having twenty-two distinct anyons, the $\mathcal{D}(D_4)$ TO is relatively simple in the sense that $D_4$ is the smallest nilpotent non-Abelian group (alongside the quaternion group), and thus $\mathcal{D}(D_4)$ is the minimal non-Abelian TO with a structure that is amenable to local decoding. While our error correction protocol is spiritually descended from that for the Toric code, the presence of non-Abelian defects in $D_4$ TO necessitates significant modifications to the previously discussed protocol. The resulting phase diagram, which we discuss in detail in Sec.~\ref{sec:phases}, is shown in Fig.~\ref{fig:2dphase}(b).


\subsection{$D_4$ Topological Order: Quasi-stabilizer model and anyons}
\label{sec:d4quasi}

\begin{figure}
\includegraphics[width=\columnwidth]{{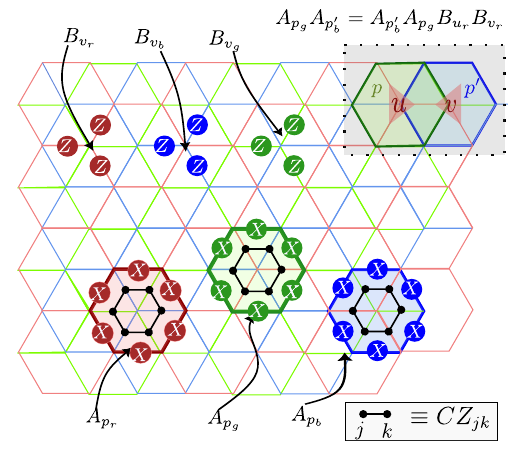}}%
\caption{\label{fig:d4-stab} Stabilizers for the $D_4$ model: qubits are located on edges of three interpenetrating honeycomb lattices, depicted in red, blue, and green. We use PBC in both vertical and horizontal directions, as defined in Eq.~\eqref{eq:tc-pbc}, for each of these lattices. A vertex operator $B_{v_c}$ on sublattice $1$ of each lattice color is shown in the upper left; the vertex operators on sublattice $2$ are products of $Z$ operators on the remaining three edges at each vertex. A plaquette operator $A_{p_c}$ for each color is shown on the bottom left. Pauli $X$ operators act on the shaded $c$ colored edges; bold black lines indicate the two-qubit control-Z gates ($CZ_{j,k}=\frac{1}{2}(1+Z_j+Z_k-Z_jZ_k)$) acting on pairs of qubits inside the plaquette. The box in the upper right shows plaquette operators that fail to commute, and instead satisfy Eq.~\eqref{eq:d4-commute-relation} with $c=\text{green},c'=\text{blue},c''=\text{red}$.}
\end{figure}

We begin with a brief overview of the properties of $D_4$ topological order, as well as the operators that can be used to stabilize $D_4$ topologically ordered ground states on the lattice in a conventional qubit architecture. $D_4$ TO describes the phase of matter obtained by considering a discrete gauge theory with a gauge group $D_4$ (i.e., the dihedral group with 8 elements) in $d=2$ spatial dimensions. The point-like excitations, or anyons, in this theory can be understood in terms of their (electric) charge and (magnetic) flux. The pure flux excitations are associated with the four non-trivial conjugacy classes of $D_4$, while pure charges are associated with the four non-trivial irreducible representations (irreps) of $D_4$. The remaining excitations, known as dyons, are labeled by a choice of conjugacy class together with a non-trivial representation of its centralizer. For $D_4$, this leads to a total of twenty-one non-trivial anyons (or twenty-two total anyons, including the identity). A more comprehensive description of this anyon theory is provided in Appendix~\ref{app:D4review}.

Since $D_4$ is a non-Abelian group, some of these anyons are non-Abelian: when a pair of such anyons is brought together (or fused), the resulting anyon type is not uniquely determined. For example, $D_4$ has one 2-dimensional irrep; two charges of this type can be combined in multiple ways, yielding different charge types (i.e., multiple irreps), in a manner analogous to the addition of angular momentum for irreps of SU(2). 

Unlike generic TOs however, the non-Abelian anyons in $\mathcal{D}(D_4)$ have fusion rules that are {\it acyclic}~\cite{galindo2018}, meaning that when a pair of non-Abelian anyons of the same type is brought together, the possible fusion channels are all Abelian anyons. This property allows $D_4$ non-Abelian states to be prepared efficiently~\cite{nathierarchy} and makes this system more amenable to conventional quantum error correction than generic non-Abelian TOs~\cite{DauphinaisPoulin}. As we will shortly show, this property is also key to our local correction protocol for heralded noise.  

In the remainder of this Section, we review a specific model Hamiltonian, introduced in Ref.~\cite{yoshida2016}, that realizes the $\mathcal{D}(D_4)$ topological phase.
This model has two advantages: first, the local Hilbert space is composed of qubits, rendering it simpler to realize on current quantum hardware~\cite{iqbal2024nonabelian}; and second, the Hamiltonian can be written as a sum of local ``check" operators, each of which individually take eigenvalue $1$ in the ground state subspace despite not commuting in the full Hilbert space. 

The model is defined on three inter-penetrating honeycomb lattices with a qubit placed on each edge, as shown in Fig.~\ref{fig:d4-stab}. For convenience, the three honeycomb lattices are shown in different colors (red, green, and blue in the Figure). 
With each color, we associate two types of check operators that stabilize the ground states. First, the vertex check centered at vertex $v_c$ is defined as
\eq{eq:d4-b-stab}{B_{v_c} = \prod_{j\in ^*{v_c}}Z_j \, ,}
where $c=r,g,b$ indicates the color of vertex $v$ and $^*v_c$ corresponds to the set of edges (of color $c$) incident on the vertex. For each color, these operators are identical to the vertex operators of the honeycomb lattice Toric code given in Eq.~\eqref{Eq:TCStabs}. Second, the plaquette checks are comprised of a product of a plaquette operator of the honeycomb lattice Toric code (of color $c$) and control-$Z$ gates acting on the enclosed edges of the remaining two honeycomb lattices (of color $c',c'' \neq c$). The plaquette operator acting on a red colored plaquette $p_r$, for instance, is given by 
\eq{eq:d4-a-stab}{A_{p_r} = \prod_{j\in ^*\mathcal{I}(p_r)} CZ_{j,j+1}\prod_{k\in\partial p_r}X_k \, .}
Here, $X$ acts on the red edges along the boundary $\partial p_r$ of the plaquette $p_r$ and $\mathcal{I}(p_r)$ represents the two vertices (one blue and one green) at the center of $p_r$. The blue and green qubits in the interior of $p_r$ are coupled via nearest-neighbor control-Z gates ($CZ$), as shown in Fig.~\ref{fig:d4-stab}. The remaining plaquette operators $A_{p_b}$ and $A_{p_g}$ are defined analogously.

Note that the plaquette check operators defined above are non-Pauli ``quasi-stabilizers". Specifically, unlike standard stabilizer operators on an $N$ qubit Hilbert space, these check operators are not elements of the $N$ qubit Pauli group, and do not commute with each other as operators.  However, they do commute within the ground-state subspace (see below for details), and stabilize the $D_4$ ground states. Thus, in the interest of brevity (and in an abuse of notation), we will often refer to $A_p$ and $B_v$ operators simply as stabilizers and will emphasize the non-trivial consequences of their ``quasi-stabilizer" nature whenever necessary.

The full Hamiltonian is given by
\eq{Eq:D4Ham}{
H = - \sum_{\text{c} = \text{r,g,b}} \left( \sum_{v_c} B_{v_c} + \sum_{p_c} A_{p_c} \right ) \, .
}
As with the Toric code, it is straightforward to verify that all vertex stabilizers commute with each other and also with the plaquette stabilizers:
\be 
[B_{v_c},B_{v'_{c'}}] = [B_{v_c}, A_{p_{c'}}] = 0 \,\, \forall \, p,v,v',c,c' \, .
\ee
However, the stabilizers on adjacent plaquettes $p_c$ and $p'_{c'}$ of different colors (see Fig.~\ref{fig:d4-stab}) do not commute; rather, 
\eq{eq:d4-commute-relation}{A_{p_c}A_{p'_{c'}}=A_{p'_{c'}}A_{p_c}B_{u_{c''}}B_{v_{c''}}}
where $u_{c''}\in\mathcal{I}(p_c)$ and $v_{c''}\in\mathcal{I}(p'_{c'})$ are vertices of the third lattice (of color $c''\neq c\neq c'$) located at the center of plaquettes $p$ and $p'$ respectively, as shown in Fig.~\ref{fig:d4-stab}. Nevertheless, Eq.~\eqref{eq:d4-commute-relation} implies that the plaquette operators do commute when acting on states for which $B_v = 1$ at every vertex. On this subset of states, one can still find simultaneous eigenstates of all the plaquette stabilizers ~\cite{tantivasadakarn2023shortest}.  
In particular, a ground state $|\psi_{D_4}\rangle$ of the Hamiltonian~\eqref{Eq:D4Ham} satisfies:
\eq{}{A_{p_c}|\psi_{D_4}\rangle=B_{v_c}|\psi_{D_4}\rangle=|\psi_{D_4}\rangle,\qquad \forall \ c,p_c,v_c \, .}
Ground states of the Hamiltonian in Eq.~\eqref{Eq:D4Ham} are thus frustration-free.

In the presence of vertex violations $B_v=-1$, on the other hand, adjacent plaquette checks fail to commute and hence cannot be simultaneously diagonalized. In particular, in the presence of a vertex defect, it is not possible for all plaquettes to be simultaneously in their ground state. An immediate consequence is that plaquette defects can effectively be absorbed by these vertex defects -- our first indication that the vertex defects are non-Abelian anyons.

Thus, unlike in the Toric code, a string of Pauli $X$ operators on the lattice has a qualitatively different effect on the ground states $|\psi_{D_4}\rangle$ than a string of Pauli $Z$ operators on the dual lattice. An error $Z_{j_b}$ on a blue edge will commute with all $B_{v_c}$, but anti-commute with a pair of adjacent blue plaquette terms that share this edge (and similarly for other colors). Dual strings of $Z$-type errors thus create pairs of Abelian plaquette defects $A_p=-1$ at their end-points; these can be pair-annihilated by re-applying the same string.  

An error $X_{j_b}$ on a blue edge will similarly anticommute with $B_{v_b}$ at the two neighboring blue vertices, thus creating a pair of vertex defects. Additionally, however, such an $X$ error fails to commute with $A_p$ on the red and green plaquettes that enclose this blue edge. It has following non-trivial relation with the $CZ$ gates present in the plaquette stabilizer:
\eq{eq:cz-relation}{X_j (CZ)_{jk} = (CZ)_{jk} X_j Z_k \, ,}
where $CZ_{jk}=\frac{1}{2}(1+Z_j+Z_k-Z_jZ_k)$ is a control-Z gate acting on qubits $j$ and $k$. This implies that the action of $X_{j_b}$ on the ground state is 
\eq{Eq:GsXact}{\eqsp{
 A_{p_r}X_{j_b}|\psi_{D_4}\rangle &= X_{j_b}Z_{k_g}Z_{k'_g} A_{p_r} |\psi_{D_4}\rangle \\
 &= X_{j_b}Z_{k_g}Z_{k'_g} |\psi_{D_4}\rangle \, ,
}}
where $A_{p_r}$ is a red plaquette that encloses the blue edge $j_b$ and $k_g, k'_g\in\ ^*\mathcal{I}(p_r)$ denote the green edges inside this red plaquette that are connected by $CZ$ gates to the edge $j_b$. 
Thus, in configurations with $Z_{k_g}Z_{k'_g} = +1$, measuring $A_{p_r}$ after the $X_{j_b}$ error acts on the ground state $|\psi_{D_4}\rangle$ will find no plaquette error, while in configurations with $Z_{k_g}Z_{k'_g} = -1$ a plaquette error will be found. Since both configurations occur in the ground state, $X_{j_b}$ leaves this red plaquette in a superposition of having and not having a plaquette defect. Similarly, acting with a longer string of $X$ operators on blue edges will result in a pair of vertex violations at the string's endpoints, together with a superposition of violated and unviolated plaquette stabilizers along the string's length, as depicted by the highlighted plaquettes in Fig.~\ref{fig:d4-x-crc}(a).

\begin{figure}
\includegraphics[width=\columnwidth]{{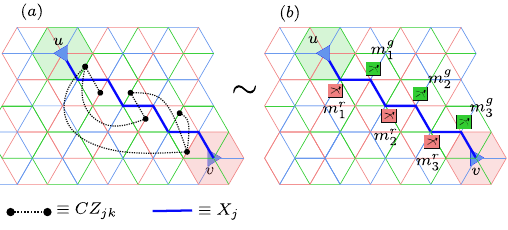}}%
\caption{\label{fig:d4-x-cz-string} Circuits for pair-creating non-Abelian vertex defects: (a) A unitary circuit $\widetilde{X}_{u,v}$ as defined in Eq.~\eqref{eq:nonabelian-cz-circuit}, where $X$ acts on bold-face blue colored edges along with a set of $CZ$ gates (shown using dotted lines) that couple red edges with all preceding green edges for red and green edges that are connected to vertices along the path. (b) A similar effect can be achieved by first applying $X$ on blue qubits as before, and then measuring the $A_{p_g}$ and $A_{p_r}$ operators indicated by green and red colored boxes, respectively. If the outcomes $m^g,m^r \equiv 1$ are post-selected on the condition of observing 
no plaquette defects, the effect is the same as applying the non-Abelian anyon string in (a). Importantly, any measurement outcome leads to a state in which re-applying the string of Pauli-$X$'s annihilates all non-Abelian defects, but leaves the system in a superposition of states with different numbers of Abelian anyons along the string's length. Both of these circuits generate $B_u=B_v=-1$ vertex defects at the endpoint of the string, with the highlighted plaquettes (centered around the endpoints) in a superposition state such that $\langle A\rangle=0$.}
\end{figure}

The underlying reason for this difference between $X$ and $Z$ errors stems from the fact that while $Z$ errors create Abelian (plaquette) defects, the vertex defects created by $X$ errors are non-Abelian anyons. By definition, the latter cannot be pair-created by a string of Pauli operators~\cite{iqbal2024nonabelian}. Instead, creating a pair of isolated non-Abelian anyons at vertices $u$ and $v$ of the same honeycomb lattice (say, blue) requires a unitary circuit with linear depth (in distance between $u$ and $v$):
\be
\label{eq:nonabelian-cz-circuit}
\widetilde{X}_{u,v} = \prod_{\ell_b \in C_b} X_{\ell_b} \prod_{\substack{\ell_r > \ell_g \\ \ell_r,\ell_g \in C_{rg}}} CZ_{\ell_r,\ell_g} \, .
\ee
Here, $C_b$ is an open string along the blue honeycomb lattice that connects vertices $u$ and $v$, and $\ell_b \in C_b$ denotes links along this path (see Fig.~\ref{fig:d4-x-cz-string}(a)). The set $C_{rg}$ consists of those links on the red and green lattices which are incident on vertices along $C_b$ and are also either enclosed by or are on the boundary of two plaquettes that enclose vertices in $C_b$\footnote{Let $\{V_c\}$ be the set of vertices along the string $C_b$ and $\{P_c\}$ be the set of plaquettes whose centers lie along $C_b$, where $c\in\{r,g\}$. Denote by $C^v_{rg}$ the set of red and green links which are incident on vertices along $C_b$ i.e., $C^v_{rg} = \{\ell_c: \ell_c \in {^*}v_{c} (c=r,g)\}$ for $v_c \in \{V_c\}$. Denote by $C^p_{rg}$ those red and green links which are shared by pairs of plaquettes along $C_b$ i.e., $C^p_{rg} = \{\ell_c: \ell_c \in \partial p_c  \, \& \, \ell_c \in \partial p'_c (c = r,g)\}$ for $p_c \neq p'_c \in \{P_c\}$. Then $C_{rg} = C^v_{rg} \cap C^p_{rg}$.}. The notation ``$\ell_r > \ell_g$" indicates those pairs of red (connected to vertices $v_r$ along $C_b$) and green (connected to vertices $v_g$ along $C_b$) links in $C_{rg}$ for which $v_r$ is to the right of $v_g$. We can similarly define open-string operators that pair-create isolated red and green vertex defects.

Using the definition of the plaquette stabilizers and the relation in Eq.~\eqref{eq:cz-relation}, it is straightforward to verify that $\widetilde{X}_{u,v}$ creates defects only at the string's endpoints: if $p$ is a plaquette that encloses a pair of edges in $C_b$, then  $[A_{p_r},\widetilde{X}_{u,v}]=[A_{p_g},\widetilde{X}_{u,v}]=0$ and hence these stabilizers remain in their ground state after applying $\widetilde{X}_{u,v}$.  Thus, plaquette defects can occur only on the two plaquettes that enclose the vertices $u$ and $v$ (for which  $\langle A_p\rangle=0$ after applying the $\widetilde{X}_{u,v}$ operator.)

In other words, to create only a pair of vertex defects, a Pauli-X string acting on blue edges must be decorated with a sequence of $CZ$ gates that couple green and red edges along the string's full length; the non-Abelian anyon string is thus inherently non-local. A bare undecorated string of $X$ operators, in contrast, cannot simply create a pair of point-like excitations; rather, it creates a veritable anyon soup (i.e. a linear combination of states with all different numbers of anyons) along the string's length; the acyclic nature of the fusion rules in the $\mathcal{D}(D_4)$ anyon theory is reflected in the fact that this soup contains only Abelian anyons. Note, however, that measuring the plaquette operators away from the string's end-points has some probability to project all of these plaquettes onto their ground states, leaving only a pair of point-like defects at the string's endpoints. In other words, these measurements, applied after the $X$ string, have some probability of effectively acting like the non-Abelian anyon string operator as shown in Fig.~\ref{fig:d4-x-cz-string}(b)\footnote{
Evidently, the intermediate plaquette operators commute with each other and can thus be measured simultaneously, as by assumption there are no blue vertex defects along the string.}. Even for general measurement outcomes, re-applying the Pauli-$X$ string after measurements does not return the stabilizers to their initial states.  Thus in a situation with active dynamics involving plaquette measurements, it is also not safe to assume that what began as a simple string of Pauli-$X$ errors acting on the ground state can be undone simply by re-applying this same string.  This is the essence of the complications that we will encounter, relative to the Toric code, in constructing our decoder below.  

The full spectrum of point-like excitations (or superselection sectors) in this model can be described in terms of combinations of the three colors of vertex and plaquette defects, yielding the twenty-two anyons of the $D_4$ TO~\cite{tantivasadakarn2023shortest}. The correspondence between the lattice defects discussed above and the conventional labels for these anyons by a conjugacy class and irreducible representation of its centralizer is summarized in Appendix~\ref{app:D4review} (see Table~\ref{tab:mapping}).

\subsection{Error Correction Protocol}
\label{sec:d4correct}
\begin{figure}
\includegraphics[width=\columnwidth]{{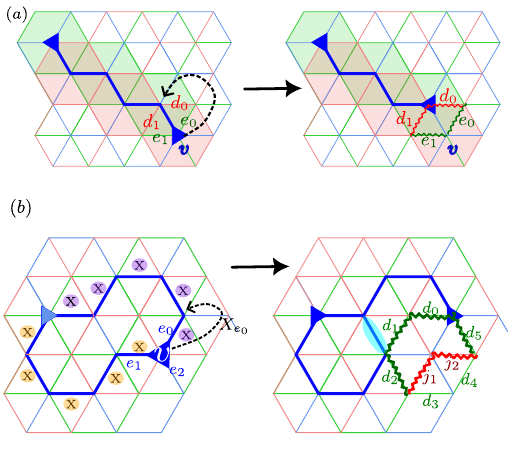}}
\caption{\label{fig:d4-x-crc} Correction protocol for $X$ errors in the $D_4$ model. (a) X-leaf move: A string of $X$ erasure errors, shown using thick blue lines, has vertex defects $B_v=-1$ at each endpoint. The highlighted plaquettes of red and green color along the length of the string are in a superposition state, as described by Eq.~\eqref{Eq:GsXact}. A blue vertex $v_b$ has a single $X$ flag on the surrounding blue edges. This flag is removed, and the vertex defect (if present) is displaced by applying $X$ to the corresponding qubit. Following this, new $Z$ flags are raised on edges $e_0,e_1,d_0$ and $d_1$ (wavy lines) to herald potential plaquette defects created during this correction step. (b) X-loop move.  Left: a configuration with $X$ flags on both edges $(e_0$ and $e_1)$ above and to the left of vertex $v$, which will result in an $X$-loop move. Here, we show a configuration that has a loop of blue $X$ flags, along with two possible $X$ error strings (colored in purple and yellow) consistent with the flags that can generate the pair of defects shown using blue triangles. Right: the loop move removes the flags on  $e_0$ and $e_1$, and displaces any vertex defects up and to the right by applying $X_{e_0}$. Wavy lines indicate the edges $\Xi[v]=\{d_0,d_1,d_2,d_5,j_1,j_2\}$ surrounding the vertex where new $Z$ flags are raised (if not already present). Additionally, as in Fig.~\ref{fig:tc-x-crc}, new $X$ flags on the outer edges of the plaquette are raised if they were not already present (in this case, a single new $X$ flag, highlighted in cyan).
} 
\end{figure}
\begin{figure*}
\includegraphics[width=\linewidth]{{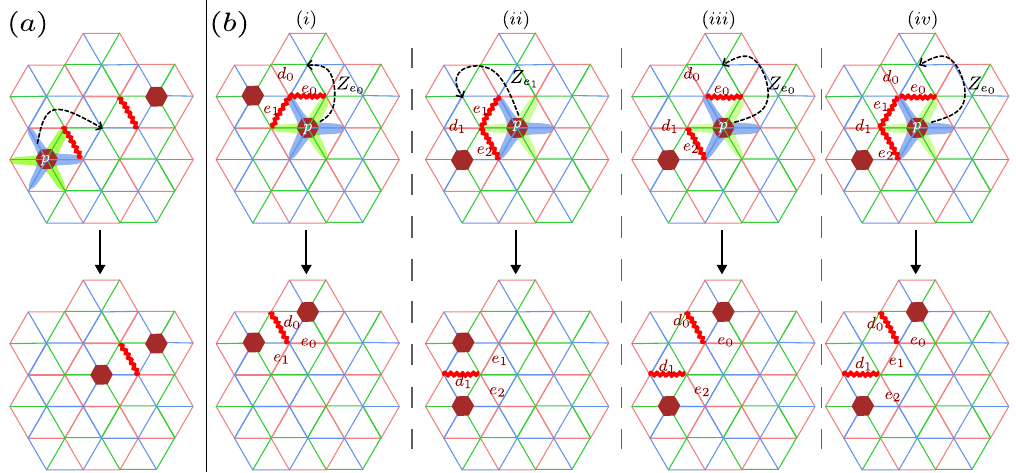}}
\caption{\label{fig:d4-z-crc} Correction protocol for $Z$ errors in the $D_4$ model. The dots indicate the plaquette defect $A_{p_r}=-1$, and wavy lines represent a $Z$ flag on the edge. The correction proceeds at a plaquette $p$ only if $X$ flags on all qubits (highlighted in green and blue) connected to the vertices $\mathcal{I}(p)$ enclosed by the plaquette $p$ are in the state $n^f_X=0$. (a) Z-leaf move: The defect on plaquette $p$ is displaced in the direction of the edge that has $Z$ flag on it. (b) Z-loop moves: Four different configurations where plaquette $p$ has $Z$ flags on edges $(e_0,e_1),(e_1,e_2),(e_0,e_2)$, and $(e_0,e_1,e_2)$ are shown in sub-panels $(i)-(iv)$. The plaquette defect is pushed towards the topmost neighbor via application of $Z$. New $Z$ flags (shown in red) are raised on the edges $d_0,d_1$ outside the plaquette, similarly to the Toric code protocol (see Fig.\ref{fig:tc-z-crc}).}
\end{figure*}

We now describe the local correction protocol that we use to stabilize a steady-state that lies in the same mixed-state phase as the ground space of the $D_4$ Hamiltonian Eq.~\eqref{Eq:D4Ham}, for noise rates below a finite threshold. Note that the heralded noise channels are the same as in the Toric code (Eqs.~\eqref{Eq:LindbladNoise},\eqref{eq:tc-noise}), aside from additional color indices. Below, the color index $c$ is suppressed for convenience wherever it does not appear explicitly.

As with the correction protocol for the 2d Toric code (see Sec.~\ref{sec:tccorrect}), our correction protocol for $D_4$ involves two types of processes. The first are leaf moves, which shrink strings of flags from their end-points inwards, while simultaneously moving any nearby stabilizer defects. The second are moves that can break open loops of flags, while pushing the corresponding stabilizer defects in a fixed direction (chosen to be the North-Western corner here) such that they necessarily remain adjacent to at least one flagged edge. In the $D_4$ case, however, a new feature arises due to the fact that the plaquette operators $A_p$ are ``decorated."

More precisely, measuring plaquette stabilizers between the application and removal of Pauli-$X$ errors can create a situation where removing the $X$ error leaves behind a plaquette defect ($Z$-error), as described in Eq.~\eqref{Eq:GsXact}. When all $X$-operators are flagged, this can be avoided by only measuring plaquette checks when none of the interior edges have $X$ flags. However, $X$ loop moves introduced in the Toric code protocol can generate configurations where the flags track the locations of previous $X$-operators only up to closed Pauli-$X$ loops; thus, we cannot be certain that correcting $X$ errors will not introduce new plaquette defects. This leads to two fundamental differences between our dynamical protocol for the $D_4$ model and its Toric code counterpart. First, the dynamics of stabilizer defects created as a result of $X$-type and $Z$-type errors (and therefore the dynamics of the $X$ and $Z$ flags) are no longer independent of each other, since we must add new $Z$ flags as we correct $X$ errors. Second, the non-trivial commutation relation of $X_j$ with the surrounding $A_p$ means that the dynamics of plaquette defects is no longer diagonal in the stabilizer basis, and the conventional stabilizer tableau method fails.

\paragraph{X flags and B defects:} We begin by discussing how the correction protocol acts on $X$ flags and the associated vertex defects. The basic idea underlying the correction protocol for $X$ flags and vertex defects is similar to that used in the Toric code. At a vertex where only one of the edges hosts an $X$ flag, we first measure $B_v$, then apply $X$ on the flagged edge if $B_v=-1$, and finally remove the flag -- exactly as we did for the Toric code. The leaf operations must be supplemented with an $X$ loop move to ensure that the protocol can remove contractible loops of flags. As noted above, however, these correction steps may inadvertently create new $Z$-errors: thus, after this correction step, we must also raise $Z$ flags to indicate the locations of the possible new $Z$-errors, as shown in Fig.~\ref{fig:d4-x-crc}.

We now describe these two types of moves in more detail. The loop moves act on the $X$ flags and $B_v$ defects exactly as for the Toric code. An example of the resulting loop moves, including the positions of the new $Z$ flags, is shown in Fig.~\ref{fig:d4-x-crc} (b). There, a vertex $v_b$ on sublattice 1 (sublattice formed by south-east corners of blue honeycombs) of the blue honeycomb lattice has $X$ flags on the blue edges above it ($e_0$) and to its left ($e_1$) and no $X$ flag on the remaining edge $e_2$, ensuring that a loop move will be applied. The loop move removes the two $X$ flags on these edges, adds $X$ flags on the remaining four edges of the plaquette (if these are not already present), and pushes the vertex defect upwards and to the right by applying $X_{e_0}$. If the defect was created by an $X_{e_0}$ error, then we have exactly undone this error. However, if the defect was generated by an $X_{e_1}$ error, this correction leads to a net application of $X_{e_0}X_{e_1}$ to the initial state, in spite of the fact that the edges $e_0$ and $e_1$ no longer carry $X$ flags. From Eq. (\ref{Eq:GsXact}), we see that subsequent measurements of plaquette stabilizers (which occur during the $Z$ correction step described later) on this state will result in
\eq{eq:cz-on-loop}{\begin{aligned}
(1+s_g A_{p_g})X_{e_0}X_{e_1}|\psi_0\rangle &= (1+s_g Z_{j_1}Z_{j_2})X_{e_0}X_{e_1}|\psi_0\rangle\\
(1+s_r A_{p_r})X_{e_0}X_{e_1}|\psi_0\rangle &= (1+s_r Z_{d_0}Z_{d_5})X_{e_0}X_{e_1}|\psi_0\rangle\\
(1+s'_r A_{p'_r})X_{e_0}X_{e_1}|\psi_0\rangle &= (1+s'_r Z_{d_1}Z_{d_2})X_{e_0}X_{e_1}|\psi_0\rangle,
\end{aligned}}
where $s_j\in\{1,-1\}$ denotes the measurement value obtained for $A_{p_j}$. Here, $A_{p_g}$ is the green plaquette enclosing the vertex $v_b$ and $A_{p_r}$, $A_{p'_r}$ are green plaquettes centered on the endpoints of the edges $e_0,e_1$ respectively. The action of $I\pm ZZ$ based on the measurement outcomes $s_g,s_r,s'_r$ can generate new defects on the adjacent plaquettes.

Instead of attempting to avoid creating such defects, our strategy is simply to ensure that they are properly heralded, allowing them to be corrected at a later stage. We therefore add $Z$ flags to each of the six qubits in $\Xi(v_c):=\{d_0,d_1,d_2,d_5,j_1,j_2\}$, thereby heralding any Abelian defects that may potentially appear as a result of this $X$ correction process. The jump operator implementing this sequence at a vertex on sublattice-1 of $c$-colored lattice can be written as
\eq{eq:d4-x-loop}{L^{{D_4}^{\text{X-loop}}}_{v_c,s,\mathbf{m^x},\mathbf{m^z}} = L^{\text{TC}_{(c)}^{\text{X-loop}}}_{v_c,s,\mathbf{m^x}} \prod_{k\in\Xi(v_c)} f^+_{Z_k}(f^-_{Z_k})^{\mathbf{m^z}_k} ,}
where $L^{\text{TC}(c)}$ is the Toric code jump operator defined in Eq.~\eqref{eq:tc-x-loop} and $\Xi(v_c)$ represents the set of six qubits in Eq.~\eqref{eq:cz-on-loop}, where new $Z$ flags are added if not already present. The bit string $\mathbf{m^z}\in\{0,1\}^{\otimes 6}$ accounts for all possible configurations of pre-existing $Z$ flags on the edges in $\Xi(v_c)$ where we recall the relation $f^+_z f^-_Z=n^f_Z$. Similar to the Toric code case, the loop moves are applied to the vertices of only one of the sublattices of each $c$-colored honeycomb lattice.

The process for adding flags after a leaf move is similar. As noted before, the vertex defects are corrected either along the path they were generated or along an alternative path that is equivalent to the original one, up to a product of stabilizers. While the former correction leaves the state unchanged, the latter results in the action of closed loops of Pauli $X$ operators. Thus, new $Z$ flags must also be added after leaf moves. Specifically, after a leaf move that results in overall application of  $X_{j_b}$, measuring $A_{p_g}$ or $A_{p_r}$ on one of the plaquettes centered on the two endpoints of the edge $e_{j_b}$ gives:
\eq{}{\eqsp{
(1+s_g A_{p_g})X_{j_b}|\psi_0\rangle &= (1+s_g Z_{d_0}Z_{d_1})X_{j_b}|\psi_0\rangle\\
(1+s_r A_{p_r})X_{j_b}|\psi_0\rangle &= (1+s_r Z_{e_0}Z_{e_1})X_{j_b}|\psi_0\rangle,
}}
After a leaf move, new $Z$ flags are therefore raised on the four edges $\Omega(j)=\{e_0,e_1,d_0,d_1\}$, as shown in Fig.~\ref{fig:d4-x-crc}(a); these ensure that any such plaquette defects are heralded. Overall, the modified $X$ leaf correction procedure is given by 
\eq{eq:d4-x-leaf}{L^{{D_4}^{\text{X-leaf}}}_{v_c,j,s,\mathbf{m^z}}=L^{\text{TC}_{(c)}^{\text{X-leaf}}}_{v_c,j,s} \prod_{k\in\Omega(j)}f^+_{Z_k}(f^-_{Z_k})^{\mathbf{m^z}_k} \, ,}
where $L^{\text{TC}(c)}$ represents the analogous Toric code jump operator centered at vertex $v_c$ defined in Eq.~\eqref{eq:tc-x-leaf} and the second term raises new $Z$ flags on qubits in the set $\Omega(j)$ if they were not present before, which is accounted for by the bit string $\mathbf{m^z}\in\{0,1\}^{\otimes 4}$ and relation $f^+_Z f^-_Z=n^f_Z$.


\paragraph{Z flags and A defects:} Recall that a $Z_{k_c}$ ($c= r,g,b)$ error on a single edge anticommutes with the two $A_{p_c}$ plaquette operators sharing that edge; thus, strings of $Z$ errors create pairs of Abelian anyons that always fuse back to the vacuum $A_{p_c}=+1$ when brought together. In the previous section, we observed that new $Z$ flags and plaquette defects emerge as a result of vertex correction processes. The flags that herald these new defects are only raised once the $X$ flags in the vicinity of the plaquette are removed. If we instead measure the plaquette before the complete set of $Z$ flags is available, the defects may get pushed in a direction incompatible with that of their paired partner. We avoid this by requiring that the $X$ flags on qubits enclosed inside a plaquette $p_c$ must be removed \textit{before} measuring $A_{p_c}$. This ensures that the potential defect $A_{p_c}=-1$ will be perfectly heralded with respect to its partner defect. 

This modified leaf-correction procedure at a plaquette $p_c$ that hosts a single $Z$ flag on one of the qubits $j$ that bounds the plaquette is described by 
\eq{eq:d4-z-leaf}{L^{{D_4}^{\text{Z-leaf}}}_{p_c,j,s}=L^{\text{TC}_{(c)}^{\text{Z-leaf}}}_{p_c,j,s}\prod_{k\in^*\mathcal{I}(p_c) }(1-n^f_{X_k}) \, ,}
where $L^{\text{TC}(c)}$ represents the corresponding Toric code jump operator in Eq.~\eqref{eq:tc-z-leaf} and 
$^*\mathcal{I}(p_c)$ contains the six edges enclosed inside the plaquette $p_c$, as shown in Fig.~\ref{fig:d4-z-crc}(a);
the second term restricts the operation to configurations where $X$ flags are absent on these edges. This must be supplemented with the corresponding loop moves implemented using
\eq{eq:d4-z-loop}{L^{{D_4}^{\text{Z-loop}}}_{p_c,s,\varepsilon,\mathbf{m^z}}=L^{\text{TC}_{(c)}^{\text{Z-loop}}}_{p_c,s,\varepsilon,\mathbf{m^z}}\prod_{k\in^*\mathcal{I}(p) }(1-n^f_{X_k}) \, .}
Here, $L^{\text{TC}(c)}$ represents the Toric code jump operator discussed in Eq.~\eqref{eq:tc-z-loop} and the second term again restricts the operation to configuration without any $X$ flags in the interior of the plaquette $p_c$.

\subsection{Simulation using Non-Pauli stabilizer tableau}
\label{sec:d4simulation}

We now describe our simulation protocol for the complete Lindbladian dynamics, which includes the heralded noise as well as the correction operators discussed above, and is given by:
\eq{Eq:D4Lindbladian}{\frac{d}{dt}\rho = \sum_{c=r,g,b}\left(\mathcal{L}_{\eta_c}(\rho) + \mathcal{L}_{C_c}(\rho)\right),}
where $\mathcal{L}_{\eta_c}$ applies heralded depolarizing noise discussed in Eq.~\eqref{eq:tc-noise} to every qubit and $\mathcal{L}_{C_c}$ describes the combined correction process 
\eq{eq:d4corr}{\begin{aligned}&\mathcal{L}_{C_c} = \gamma_x \sum_{\substack{v_c,s=\pm1\\j\in^*v_c,\\ \mathbf{m^z}\in\{0,1\}^{\otimes4}}} \mathcal{L}^{{D_4}^{\text{X-leaf}}}_{v_c,j,s,\mathbf{m^z}}  \ + \ \gamma_z \sum_{\substack{p_c,s=\pm1\\ j\in\partial p_c}} \mathcal{L}^{{D_4}^{\text{Z-leaf}}}_{p_c,j,s}     \\
&+ \gamma_x \sum_{\substack{v_c\in\text{sublattice-1},\\ s=\pm1,\\ \mathbf{m^x}\in\{0,1\}^{\otimes4}\\ ,\mathbf{m^z}\in\{0,1\}^{\otimes6}}} \mathcal{L}^{{D_4}^{\text{X-loop}}}_{v_c,s,\mathbf{m^x},\mathbf{m^z}} +\gamma_z\sum_{\substack{p_c,\\ s=\pm1,\\ \varepsilon\in\mathcal{E}(p_c),\\ \mathbf{m^z}}}\mathcal{L}^{{D_4}^{\text{Z-loop}}}_{p_c,s,\varepsilon,\mathbf{m^z}} .
\end{aligned}}
Here individual Lindblad superoperators $\gamma_{\Phi}\mathcal{L}_{\Phi}=L_{\Phi}\rho L_{\Phi}^\dag-\frac{1}{2}\{L^\dag_{\Phi}L_{\Phi},\rho\}$ are defined in terms of the jump operators detailed in Eqs.~\eqref{eq:d4-x-loop},~\eqref{eq:d4-x-leaf},~\eqref{eq:d4-z-leaf}, and ~\eqref{eq:d4-z-loop}.

In our protocol, the flags undergo classical dynamics and can thus be simulated using standard Monte Carlo techniques (see \cite{hinrichsenNonequilibriumCriticalPhenomena2000}). However, as mentioned before, one complication of the above correction protocol, relative to that required for the Toric code, is that the dynamics is no longer diagonal in the quasi-stabilizer basis. This is because applying Pauli-$X$ operators creates a superposition of different configurations of Abelian anyons. Hence, simulating the full dynamics of the quantum system requires additional techniques. Here, we adapt the stabilizer tableau method~\cite{gottesman1998heisenberg,aaronson2004improved} and describe how an appropriate generalization allows us to simulate the dynamics of the quasi-stabilizer defects. Specifically, this method allows us to simulate individual measurement trajectories of the quasi-stabilizers, which we then average over independently generated measurement trajectories using a Monte Carlo approach to compute physical observables (see Appendix.~\ref{app:numerics}). The key difference in our approach, relative to usual stabilizer tableaux, is that the quasi-stabilizers defining the $D_4$ state are \textit{non-commuting} operators that are not elements of the Pauli group. We outline how this complication can be treated here, with further details provided in Appendix~\ref{StabApp}. As before, we slightly abuse the notation and refer to $A_p$ and $B_v$ as stabilizers whenever it is clear from the context.

Our system begins in a state stabilized by the $A_p$ and $B_v$ operators (the color index $c$ is implicit here), satisfying
\eq{}{A_p|\psi_0\rangle = B_v|\psi_0\rangle = |\psi_0\rangle, \,\, \forall v,p \,}
(see Eqs.~\eqref{eq:d4-b-stab},~\eqref{eq:d4-a-stab}) as well as $n^f_{X_j}=n^f_{Z_j}=0$, for all edges $j$. The entire noise and correction dynamics involves the action of the Pauli operators $X,Z$, and measurements of $A_p$ and $B_v$. The basic approach that we will use is to track how the application of Pauli operators and measurements modifies the stabilizers over the course of the simulation. In other words, we track individual trajectories in the time evolution of the qubits by time evolving these stabilizers. Since the flags undergo classical dynamics that remains diagonal in their population basis, this is sufficient to allow us to simulate individual trajectories of the full system.  

\begin{figure}
\subfloat{%
  \includegraphics[width=\columnwidth]{{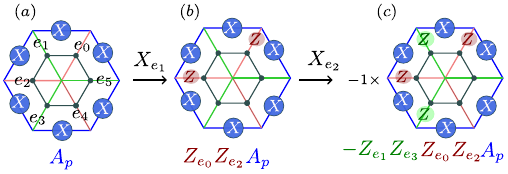}}%
}\hfill
\caption{ The modification of plaquette stabilizer $A_p$ after repeated application of $X$ on its interior edges. (a$\rightarrow$ b): applying $X$ to a green edge $e_1$ appends the product $Z_{e_0} Z_{e_2}$ of Pauli $Z$ operators on the two adjacent red edges to the original plaquette operator.  (b$\rightarrow$ c): applying $X$ to a red edge $e_2$ appends $Z_{e_1} Z_{e_3}$ to the plaquette operator along with a minus sign, due to anticommutation with the existing $Z_{e_2}$ from the previous step. Updating vertex stabilizers, or updating plaquette stabilizers after applying $Z$ to an external edge, changes the resulting stabilizers only up to a sign.  
}\label{fig:d4-tab-x}
\end{figure}

To determine the effect of applying a unitary operator, we use the fact that if $S|\psi\rangle=|\psi\rangle$ then $U S U^{-1} U |\psi\rangle=U|\psi\rangle$; thus, the stabilizer of the new state $|\psi\rangle':=U|\psi\rangle$ is given by $S'=U S U^{-1}$. Applying Pauli $Z_k$ hence gives:
\eq{eq:z-stab-update}{B_v\xrightarrow{Z_k} B_v , \qquad \pm A_p \xrightarrow{Z_k} \mp A_p \, ,}
where $k$ is one of the outer edges of plaquette $p$. Similarly, applying Pauli $X_k$ transforms the vertex operators as  
\eq{eq:x-Bstab-update}{\pm B_v\xrightarrow{X_k} \mp B_v \, ,}
where $v$ represents vertices (of same color as edge $k$) at either endpoints of edge $k$.

While applying $X$ only updates the vertex stabilizers up to a sign,  it modifies the form of the plaquette stabilizers. For example, applying $X_{e_1}$ updates the plaquette stabilizers as
\eq{eq:x-Astab-update}{A_p \xrightarrow{X_{e_1}} A'_p:= Z_{e_0}Z_{e_2} A_p }
for any plaquette $p$ that encloses the edge $e_1$, with $e_0,e_2$ being the adjacent edges (of a different color) that share the central vertex, as shown in Fig.~\ref{fig:d4-tab-x}. Here, we have used the relation $X_j CZ_{j,k} = CZ_{j,k} X_j Z_k$ and the locations of $CZ$ gates in $A_p$ to determine the modified plaquette stabilizer. As with the original plaquette operators, note that while the new plaquette terms do not commute as operators, by construction they commute when acting on the modified state $|\psi'\rangle = X_{e_1}|\psi\rangle$.

Next, to determine the effect of measurements, we use the fact that at any point during the simulation, the instantaneous stabilizers have the form $\{ \pm B_v, \pm  \prod_{k,p} Z_k A_p \}$. These operators either commute or anticommute with the operators $B_v$ and $A_p$ (when acting on the state at that instant of time) that we measure, so standard stabilizer tableau techniques can be used to simulate these measurement processes. $B_v$ commutes with all instantaneous stabilizers, and always has a definite measurement outcome given by the sign of the associated instantaneous vertex stabilizer. $A_p$, on the other hand, may either commute or anti-commute with a given instantaneous stabilizer. When it fails to commute, the instantaneous stabilizers must be modified in a manner consistent with the probabilistic measurement outcome (see Appendix~\ref{StabApp}). If $A_p$ commutes with all instantaneous stabilizers, the measurement outcome is deterministic, and the set of stabilizers need not be modified. However, it can be the case that $A_p$ {\it commutes with} all instantaneous stabilizers, but is {\it not among} the instantaneous list of stabilizer generators. In this case, this deterministic measurement outcome of $A_p$ must be determined from the existing stabilizer generators; we describe how to do this in Appendix~\ref{StabApp} and also provide additional details about encoding the stabilizers in tableaux that can be efficiently simulated.

\begin{figure*}
\includegraphics[width=\textwidth]{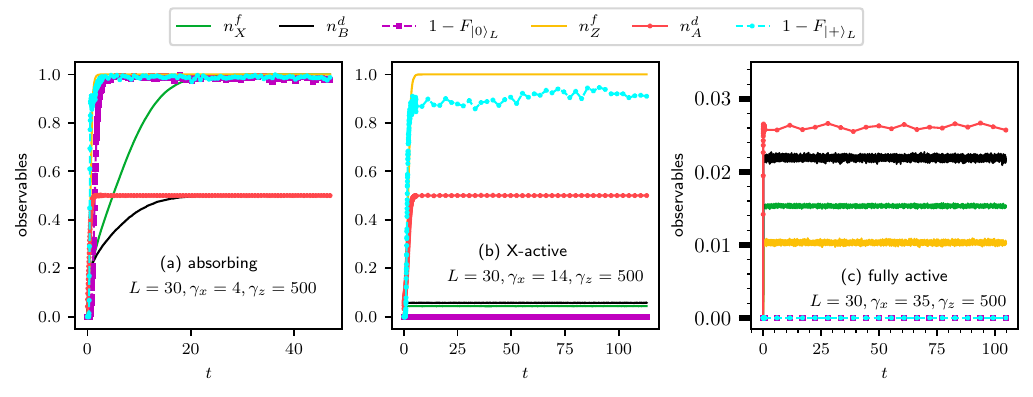}

\caption{\label{fig:d4-densityVst} Time
evolution for the $D_4$ model on three $30\times 30$ sized interpenetrating honeycomb lattices at three representative points in the $\gamma_x - \gamma_z$ phase diagram (see Fig.~\ref{fig:2dphase}(b)). Plots show the densities  of $X$ and $Z$ flags, stabilizer defect densities $n^d_B$ and $n^d_A$, and the logical error rate $1-F_{|\widetilde{0}\rangle_L(|\widetilde{+}\rangle_L)}$ (see Eq.~\eqref{Eq:LogFid}) as a function of time. Time is measured in units of inverse erasure noise rate, set to $\eta=1$. The flags are initialized to $n^f_X=n^f_Z=0$ and the initial state of qubits is set to $|\widetilde{0}\rangle_L$, except when calculating $1-F_{|\widetilde{+}\rangle_L}$, where it is initialized in the $|\widetilde{+}\rangle_L$ state. Each data point is obtained by averaging over $500$ independent Monte Carlo trajectories. (a) Absorbing phase ($\gamma_x=4,\gamma_z=500$): when the $X$ correction rate is small, $X$ flags proliferate, which also renders $Z$ correction ineffective (see Eqs.~\eqref{eq:d4-z-leaf}, \eqref{eq:d4-z-loop}). Since all flags are trapped in the absorbing state, the defect densities saturate to $1/2$, and all logical information is lost, as witnessed by $F\rightarrow 0$. (b) Partially $X$-active phase ($\gamma_x=14,\gamma_z=500$): while the densities of $X$ flags and vertex defects saturate to small values, $Z$ flags still proliferate. This phase can only protect classical information, as seen by perfect recovery of $|\widetilde{0}\rangle_L$ state (magenta curve) but failure to preserve $|\widetilde{+}\rangle_L$ state (cyan curve). (c) Active phase ($\gamma_x=35,\gamma_z=500$): when the $X$ correction rate is sufficiently large, the long-time value of $n^f_X$ (and of $n^d_B$) becomes small, allowing $Z$ correction to become sufficiently effective to stabilize  $n^f_Z$ and $n^d_A$ densities at small values. In this phase, the logical error probability is close to $0$, witnessed by $(1-F)\approx0$ for both $|\widetilde{0}\rangle_L$ and $|\widetilde{+}\rangle_L$ initial states leading to protected quantum information up to long times. }
\end{figure*}

\section{$D_4$ Steady-State Phase Diagram
}
\label{sec:phases}

Having established the local correction protocols, we now present the steady-state phase diagram under the continuous action of correction and heralded noise (see Eq.~\eqref{Eq:D4Lindbladian}). After discussing the flag dynamics and reviewing the diagnostics used to characterize the steady-state order, we present the complete dynamical phase diagram as well as the nature of the transition. Towards the end of this Section, we also briefly discuss the dynamics in the presence of imperfectly heralded noise.


\subsection{Dynamical phase diagram of flags}

We begin by discussing the steady-state phases of the flags, which undergo a classical dynamics that does not depend on the dynamics of the stabilizer defects. Recall that the active phase of the flags refers to the scenario where the absorbing state is not reached for times that are exponentially long in the system size; in contrast, regimes where the absorbing state $n^f_X= n^f_Z=1$ is the only possible choice of steady-state (i.e., where the absorbing state is reached on a time-scale that is independent of system-size) lie within the \textit{absorbing phase} of the dynamics.

Fig.~\ref{fig:2dphase}(b) shows the numerically obtained steady-state flag density as a function of the correction rates $\gamma_x$ and $\gamma_z$ (for noise rate $\eta = 1$). When both correction rates are sufficiently high, we observe a fully active phase, in which the densities of both $X$ and $Z$ flags remain low out to times that are exponentially long. On the other hand, when both correction rates are low, the steady-state belongs to the absorbing phase where all flags are raised, such that no corrections can take place. These two phases are separated by a partially active phase, in which the density of $X$ flags remains low, but $Z$ flags proliferate and rapidly reach a density $n^f_Z = 1$.  

To understand this phase diagram, we begin with the $X$ flags. It is straightforward to see that these undergo dynamics identical to those of the $X$ flags in the Toric code, since neither the presence of $Z$ flags nor the $Z$ correction steps affect the procedure used to apply $X$ corrections. In particular, the $X$ flags enter into an absorbing state precisely when the $X$ correction rate is below the threshold value for the Toric code, i.e. $\gamma_x\le\gamma^{c}_{\text{TC-X}}$. For $\gamma_x > \gamma^{c}_{\text{TC-X}}$, the $X$ flags remain in their active state.  

The dynamics of the $Z$ flags, however, is rather different compared to their behavior in the Toric code. There are two reasons behind this: first, the correction for the $Z$ flags (see Eqs.~\eqref{eq:d4-z-leaf}, \eqref{eq:d4-z-loop}) can only be carried out when all $X$ flags inside a given plaquette are lowered (i.e., in the $n^f_X=0$ state). Once the $X$ flags enter their absorbing state, the $Z$ correction protocol becomes inert and $Z$ flags also necessarily enter their absorbing state, irrespective of the correction value $\gamma_z$. Thus, when $\gamma_x\le\gamma^{c}_{\text{TC-X}}$ the system is in the fully absorbing phase irrespective of $\gamma_z$. This contrasts sharply with the case of the Toric code, where we observed a partially active phase where $X$ flags have proliferated but $Z$ flags have not.  

If the rate $\gamma_x$ is increased beyond $\gamma^{c}_{\text{TC-X}}$, the $X$ flags enter an active state. For small values of $\gamma_z$, the $Z$ flags remain in the absorbing state in both the Toric code and $D_4$ models, resulting in a partially active phase. However, the threshold value of $\gamma_z$ required to pass from this partially active phase to the fully active phase is much higher here than in the Toric code, and also depends on the value of $\gamma_x$. This is due to two key differences between the dynamics of $Z$ flags in the $D_4$ model relative to the Toric code: first, even a low density of $X$ flags can significantly lower the actual rate at which $Z$ corrections are applied, and second, removing $X$ flags introduces new $Z$ flags in the $D_4$ setup. In Appendix~\ref{app:mf}, we also provide a mean-field phase diagram, which is obtained by ignoring all inter-site flag correlations: this approximation illustrates the impact of the correlations between the locations of the $X$ and $Z$ erasures, which is absent for the Toric code dynamics, on the phase diagram.

Fig.~\ref{fig:d4-densityVst} shows the combined evolution of flag densities, stabilizer defect densities,  and logical fidelity at a representative parameter value within each distinct steady-state phase shown in Fig.~\ref{fig:2dphase}(b).  We see that in each case, the dynamics of stabilizer defects closely tracks that of the flags, with both steady-state values and the time-scales at which these reach their long-time values closely correlated. Here, the densities of the vertex and plaquette stabilizer defects are defined, respectively, as
\be
n_B^d = \frac{1}{2L^2} \sum_v \frac{\left(1 - \braket{B_v} \right)}{2} \, , \,\, n_A^d = \frac{1}{L^2} \sum_p \frac{\left(1 - \braket{A_p} \right)}{2} \, .
\ee
where summation runs over all vertices $v$ and plaquettes $p$. Fig.~\ref{fig:d4-densityVst}(a) shows a representative point in the absorbing state, where both flag densities increase rapidly to their steady-state value of $1$ (even though this increase occurs more slowly for $X$ flags, as is apparent in the figure). The densities of plaquette defects (created by $Z$ errors) and vertex defects (created by $X$ errors) closely track the respective flag densities and rapidly equilibrate to their steady-state values of $1/2$, indicating that the qubits quickly reach the maximally mixed state.

In the partially active phase, shown in Fig.~\ref{fig:d4-densityVst}(b),  $Z$ flags proliferate rapidly and attain a density of $1$, but $X$ flags remain at low density in the steady-state. Consequently, the density of plaquette stabilizer defects (caused by $Z$-type error strings) rapidly saturates to the maximally mixed value $n^d_A=1/2$, but the density of vertex stabilizer defects remains low in the steady-state (indicating a low density of $X$-type error strings).

Finally, the time evolution for the system in the fully active phase is shown in Fig.~\ref{fig:d4-densityVst}(c). Here, we see that both flag densities rapidly approach their steady-state values, but -- in contrast with the absorbing state -- these are now small. As a result, the density of stabilizer defects is also small in the steady-state.  
This low density of stabilizer defects strongly suggests that the steady-state has $D_4$ topological order. In the following Section, we will show explicitly that this is indeed the case.


\begin{figure}
  \includegraphics[width=\columnwidth]{{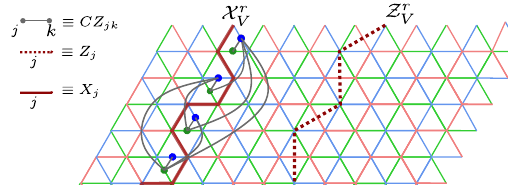}}%
\caption{\label{fig:d4-logicals}Logical operators for the $D_4$ ground space: A $12\times 6$ sized patch of $D_4$ model with PBC is shown. The $Z$ logical operator $\mathcal{Z}^r_V$ acts along a non-contractible loop on the red dual lattice in the vertical direction. The logical $X$ operator $\mathcal{X}^r_V$ is constructed by applying $X$ along a non-contractible loop $C_r$ on the red honeycomb lattice, followed by a set of $CZ$ gates that couple all blue and green edges in a specific pattern (see Eq.~\eqref{eq:nonabelian-cz-circuit}). Here, starting from the top, every blue edge that is incident to a vertex along $C_r$ is connected to all green edges below it along the path $C_r$. The remaining logical operators can be defined analogously.}
\end{figure}


\subsection{Diagnostics for topologically ordered steady-states}
\label{Sec:D4Diagonstics}

Before we present our numerical results for the dynamics of stabilizer defects, we review the diagnostics that we will use to characterize the steady-state order. We will use two metrics here. The first metric can be viewed as a pragmatic one: we ask whether, after applying a conventional MWPM error-correction protocol, our steady-state returns to the original $D_4$ topological ground state in which it was prepared. The second metric asks whether our steady-state can be connected to the defect-free $D_4$ ground state via a finite depth quasi-local quantum channel -- an affirmative answer would place the steady-state in the same mixed-state phase as the pure $D_4$ ground state~\cite{sang2023mixed}.  

Let us begin with the error-correction metric. The simultaneous $+1$ eigenspace of the $A_p$ and $B_v$ operators of the $D_4$ model is $22$-fold degenerate on the torus, reflecting the number of distinct superselection sectors (anyons) of $D_4$ TO. The states in this subspace can be distinguished based on the values of certain non-contractible loop operators. One choice is to label the $22$ ground states according to their eigenvalues under the non-contractible Abelian loop operators, given by $\mathcal{Z}^c_{H/V}=\prod_{k\in l_{H/V}}Z_k=\pm1$ where $l_H(l_V)$ are horizontal (vertical) loops on the dual of the $c$ colored honeycomb lattice, as shown in Fig~\ref{fig:d4-logicals}. We term these the ``logical-$Z$" operators. While there are $2^6$ different choices for the eigenvalues of these logical operators, only $22$ of them are consistent with $A_p=B_v=1$ for all $p,v$~\cite{iqbal2024nonabelian}.  The operators that switch between the allowed logical-$Z$ eigenstates are the non-contractible string operators associated with transporting non-Abelian anyons around the torus. We refer to these as the ``logical-$X$" operators, $\mathcal{X}^c_H,\mathcal{X}^c_V$. As shown in Fig~\ref{fig:d4-logicals}, these non-Abelian string operators consist of a string of $X$ operators decorated with non-local $CZ$ gates. A second basis choice for the $22$ locally indistinguishable ground states is by their values under the vertical $\mathcal{Z}_V$ and $\mathcal{X}_V$ logical operators. 

In order to analyze the fate of the encoded logical information in the ground-space of the $D_4$ model under the noise and correction dynamics, it will be useful to consider two different initial logical states, one in each of these basis choices: 
\eq{eq:d4-initial-state}{\eqsp{|\widetilde{0}\rangle_L&:=\bigotimes_{p_c,v_c,c}|A_{p_c}=B_{v_c}=+1;\ \mathcal{Z}_H^c=\mathcal{Z}^c_V=+1\rangle \, , \\ |\widetilde{+}\rangle_L&:=\bigotimes_{p_c,v_c,c}|A_{p_c}=B_{v_c}=+1;\ \mathcal{X}^c_V=\mathcal{Z}^c_V=+1\rangle \, .
}}
The ability to recover \emph{both} of these states from the time-evolved noisy-mixed state is a measure indicating that both the amplitude and the relative phase amongst the logical-computational states can be protected.

To determine whether this logical information is preserved in the steady state, we use a MWPM decoder appropriate to the $D_4$ TO. At any given instant, the state of the system is represented by a product of vertex eigenstates and, potentially, a superpositions of plaquette eigenstates. The decoder acts by first measuring all vertex stabilizers, and pairing up the vertex defects using minimum-weight perfect matching. Pairs of vertex defects are eliminated by applying strings of Pauli-$X$ operators to obtain a state with $B_{v}=+1$ at every vertex. As described above, this can create further plaquette $(A_p$) defects. Since all $A_p$ operators commute in the $B_{v}\equiv +1$ state, we can next measure all plaquette operators, collapsing the state onto a fixed configuration of plaquette defects. The plaquette defects are then paired up using a MWPM decoder on the dual lattice, and eliminated by applying Pauli-$Z$ operators along the appropriate dual strings. We compute the failure of this decoding process by evaluating the fidelity of the resulting decoded state $|\psi^{(r)}_{\text{decoded}}\rangle$ with the initial state: 
\eq{Eq:LogFid}{1-F_{\Phi} = 1-\frac{1}{N_r}\sum_{r=1}^{N_r}|\langle\psi^{(r)}_{\text{decoded}}\ |\ \Phi\rangle|^2 \, ,}
where $\Phi$ is one of the initial code states defined in Eq.~\eqref{eq:d4-initial-state} and the overlap is averaged over $N_r$ independent Monte Carlo realizations.

Our second diagnostic utilizes the two-way channel connectivity equivalence relation placed on the space of mixed-states~\cite{sang2023mixed}: two states $\rho_1$ and $\rho_2$ belong to the same mixed-state phase if there exist a pair of quasi-local, finite-depth quantum channels $\Sigma_{12}, \Sigma_{21}$ such that $\Sigma_{21}(\rho_1) = \rho_2$ and $\Sigma_{12}(\rho_2) = \rho_1$. In our case, the challenge is to determine whether there exists a finite-depth quantum channel that takes the steady-state density matrix $\rho_{ss}$ back to the initial error-free state $\rho_1$. We assess this by measuring the time required for the steady-state to evolve back to the error-free state under the Lindbladian Eq.~\eqref{Eq:D4Lindbladian} in the absence of noise i.e., under the action of only the error correction terms. 
Specifically, we examine how the time $t$ at which $\rho_{1} \sim e^{\mathcal{L}_{\eta=0} t}\rho_{ss}$ scales with system size, with $\mathcal{L}_{\eta=0}$ given by Eq.~\eqref{Eq:D4Lindbladian} with $\eta = 0$. The density matrices $\rho_1$ and $\rho_{ss}$ are considered as being connected via a finite-depth quantum channel as long as this time grows at most (poly)-logarithmically with $L$~\cite{sang2023mixed}. Here, two mixed states are considered to be close in terms of the fidelity, i.e. $\rho\sim\rho'$, if $F(\rho,\rho')>1-\varepsilon$ for some $\varepsilon>0$.

We emphasize that this is, in principle, a different (and stronger) criterion than requiring that the logical information encoded in the initial state can be recovered, since our MWPM decoder is highly non-local and need not guarantee the existence of a pair of quasi-local finite-depth two-way channels. Interestingly, however, at least in the context of the Toric code, Ref.~\cite{sang2023mixed} showed that the two criteria are equivalent in practice. 


\subsection{Phase diagram of the $D_4$ model}
\label{sec:D4phase}

We will now apply the criteria outlined in the previous subsection to show that, in the $D_4$ model with perfectly heralded errors, (a) the flag dynamics fully dictates the steady-state order as measured by MWPM decodability, and that (b) the active phase has non-Abelian steady-state topological order as measured by the criterion of two-way connectivity. 

We begin by discussing the correspondence between flag dynamics and the logical fidelity (\ref{Eq:LogFid}) shown in Fig.~\ref{fig:d4-densityVst} (also see Fig.~\ref{fig:2dphase}(b)). 
When both $X$ and $Z$ flags are in the absorbing state, as noted above, the 
densities of both plaquette and vertex defects closely track the respective flag densities and rapidly equilibrate to their steady-state values of $1/2$, indicating a fully decohered mixed-state. This proliferation of stabilizer defects leads to a rapid decay of the logical fidelities for both the $|\widetilde{0}\rangle_L$ and $|\widetilde{+}\rangle_L$ states.

In the fully active phase, Fig.~\ref{fig:d4-densityVst}(c), where the density of both flags and stabilizer defects is small in the steady-state, we expect that the strings of errors that create stabilizer defects (which must lie along strings of flags) have a very high probability of being short. This is clearly borne out in our numerics, as seen in Fig.~\ref{fig:d4-densityVst}(c): the probability of decoding to the incorrect logical state is vanishingly small (i.e., the logical fidelity is essentially $1$ after decoding the steady-state, for all times shown). In other words, from the perspective of decodability, our fully active phase has non-Abelian $D_4$ topological order.  

Fig.~\ref{fig:d4-densityVst}(b) shows results for the partially active regime, where the steady-state density of plaquette stabilizer defects is $1/2$, but the density of vertex stabilizer defects remains low in the steady-state (indicating a low density of $X$-type error strings). In this phase, we see a distinction between the fidelities of the MWPM-decoded state and the initial state, depending on which of the two logical states we use. Specifically, because vertex defects are rare, the probability of acting with a non-contractible loop of $X$ strings (and hence a non-contractible non-Abelian loop operator, after the full error correction protocol is executed) during the MWPM decoding step is small. This leads to a very small probability of the logical errors in $|\widetilde{0}\rangle_L$ state that is sensitive only to long strings involving $X$ operators. Logical $X$ information, on the other hand, is sensitive both to non-Abelian strings and the Abelian $Z$ strings; in this partially active phase, where strings of $Z$ errors proliferate, this information is rapidly lost. In other words, the partially active phase is a good classical memory, similar to the partially active phase in the Toric code, but fails to preserve the full quantum information of the $D_4$ state. 

\begin{figure}
 \includegraphics[width=\columnwidth]{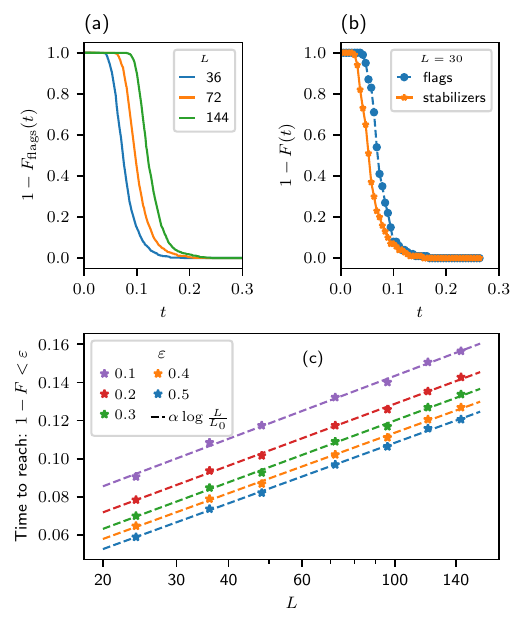}
\caption{\label{fig:d4-two-way-fid}Two way connectivity in the $D_4$ active phase: We estimate the time required for noiseless dynamics $\eta=0$ to take the (quasi)-steady-state $\rho$ in the active phase ($\gamma_x=35,\gamma_z=500$ simulated upto $500$ MC sweeps) back to the pure $D_4$ ground state $|\psi_0\rangle$. To numerically access the larger system sizes, we estimate the fidelity of the time-evolved state based only on the flag densities using $F_{\text{flags}}$ (see Eq.~\eqref{eq:flag-fid}). 
(a) $1-F_{\text{flags}}(t)$ evaluated for various system sizes with linear dimension $L$. The data is obtained by averaging over $1000$ independent Monte Carlo realizations. (b) $1-F_{\text{flags}}(t)$ (blue) and $1-F_{\text{full}}(t)$ of the qubit system (orange) are plotted as a function of time for $L=30$. This demonstrates that $F_{\text{flags}}(t)<F_{\text{full}}(t)$ and hence recovery time-scales obtained using $F_{\text{flags}}$ upperbound the recovery time computed in terms of the full fidelity of the stabilizer system. (c) Time $\tau$ to achieve $1-F_{\text{flag}}(t)<\varepsilon$ is plotted as a function of system size $L$ for various cutoff values $\varepsilon$ on a linear-log scale plot. The dotted lines show best fits to the functional form $\alpha \log{(L/L_0)}$. We note that the presence of even a single flag in a given trajectory leads to zero fidelity with $|n^f_{X,Z=0}\rangle$ state. }
\end{figure}

Next, we turn to our second criterion for examining the phase diagram, based on the existence of two-way quantum channel connectivity between the initial state and the steady-state. Fig.~\ref{fig:d4-two-way-fid} shows the fidelity $F(t) =\langle \psi_{D_4}|e^{\mathcal{L}_{\eta=0}t}[\rho_{ss}]|\psi_{D_4}\rangle$ of the  initial state  with the time-evolved steady-state $\rho_{ss}$ for system sizes ranging from $24 \times 24$ sites to $144 \times 144$ sites. (Here $\mathcal{L}_{\eta=0}$ indicates time evolution under the correction-only Lindbladian).  To allow simulations out to larger system sizes, we simulate only the flag dynamics, and plot the quantity
\eq{eq:flag-fid}{F_{\text{flag}}(t):=\langle n^f_X=n^f_Z=0|\rho_{\text{flag}}(t)|n^f_X=0,n^f_Z=0\rangle}
which estimates the logical fidelity at time $t$ from the flag density. Here, $|n^f_{X,Z}=0\rangle$ is state where both $X$ and $Z$ flags on all sites are set to $0$ and $\rho_{\text{flag}}$ is the reduced density matrix of the flags. This upper bounds the full fidelity, since a state with no flags automatically has no defects or error strings, and thus will error-correct into the correct logical state. We verify in Fig.~\ref{fig:d4-two-way-fid}(b) that for $L\le30$, where we can perform the full stabilizer simulation, this is a good approximation for the full fidelity. 

As seen in Fig.~\ref{fig:d4-two-way-fid} (c), the resulting fidelities are consistent with a local correction time that scales logarithmically with the system size, providing compelling evidence that our steady-state is in the same mixed-state phase as the pure $D_4$ topologically ordered state. While the time-scale for the state to get close to the initial flag-free state in terms of fidelity increases logarithmically with increasing $L$, the local densities of flags approach zero in system-size independent time. Since correction can only remove flags from the boundaries of flag clusters, the timescale to recover the initial state should be determined by the size of the largest connected clusters of flags. In Appendix~\ref{app:cluster}, we show that this is indeed the case: we plot the distribution of flag-clusters as a function of their size, and find that the average size of the largest cluster increases logarithmically with increasing system size. We also expect that the partially active phase we have found here represents an intrinsically mixed-state TO (distinct from the imTO characterizing the active phases of the Toric code) since it encodes a non-trivial classical memory (see Refs.~\cite{sohal2024imto,zhang2024} for a discussion of such intrinsically mixed TO phases).

\begin{figure}
\includegraphics[width=\columnwidth]{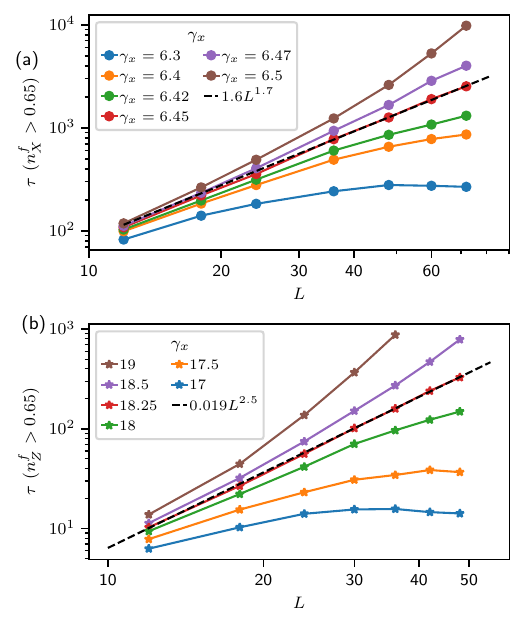}

\caption{\label{fig:first-order-tau}Determination of first order transition point based on the decay of flag densities towards the absorbing state in a finite-sized system: the time $\tau$ required for the erasure flag density to cross a finite constant value, chosen to be $n^f_0=0.65<1$, when initialized in a state with no flags, is plotted as a function of linear dimension $L$ for varying values of $X$ correction rates $\gamma_x$ close to the transition. The critical transition rate is characterized by the value of $\gamma_x$ where the curve changes from concave up (active phase for infinite system) to concave down (absorbing phase for infinite system). (a) The time $\tau$ for $n^f_X$ to reach $0.65$ as a function of system size for parameters near the transition from absorbing to partially X-active (where $Z$ flags are in their absorbing state) phase. Because the $X$ flags on a given honeycomb lattice evolve independently of the $Z$ flags (and the $X$ flags of the remaining two lattices), we only simulate $X$ flags of a single honeycomb lattice, allowing access to larger system sizes. Each data point is obtained by averaging over $1000$ MC realizations. (b) $\tau$ associated with the $Z$ flag density is shown as a function of system size for  $\gamma_z=500$ and varying values of $\gamma_x$ close to the transition from partially X-active to fully active phase. The data-points are obtained by averaging over 500 MC realizations.}
\end{figure}

\begin{figure}
\subfloat{%
  \includegraphics[width=\columnwidth]{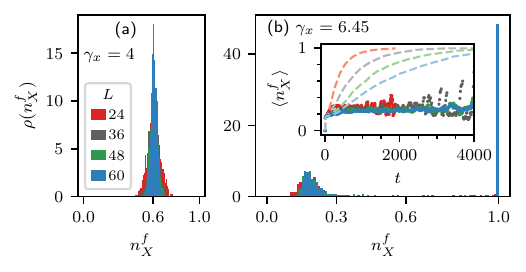}
}%

\subfloat{%
  \includegraphics[width=\columnwidth]{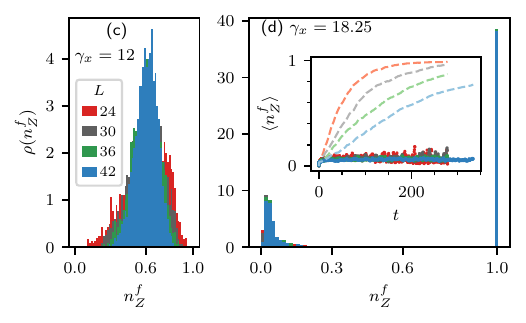}
}
\caption{\label{fig:x-flag-hist} Bistability at the transitions: the probability distribution function $\rho(n^f_{X(Z)})$ of the density of the number of $X(Z)$ flags ($n^f_{X(Z)})$ among Monte Carlo trajectories at time $t_*$. The time $t_*$ is chosen such that the average density reaches an arbitrary finite value $n^f_{\text{avg}}=0.6$. The data is shown for $500-2000$ independent Monte Carlo realizations. (a-b) The transition from absorbing to partially active state is analyzed by monitoring the density of $X$ flags on one of the honeycomb lattices of the $D_4$ model. (a) Absorbing phase (shown for $\gamma_x=4$, independent of $\gamma_z$): the distribution has a single peak centered near the average value and it moves toward $n^f_X=1$. (b) Near the absorbing to partially active transition ($\gamma_x=6.45$, independent of $\gamma_z$): the distribution is bimodal where some of the Monte Carlo trajectories enter the absorbing state ($n^f_X=1$) but a finite fraction of trajectories remain active with a small density of $X$ flags, witnessed by a distinct secondary peak at $n^f_X<0.5$. The inset shows the full $n^f_X$ (dashed lines) and the average density of flags conditioned on Monte Carlo realizations still in the active state (solid dots) as a function of time. (c-d) Similar results are shown for the transition from $X$-active to fully active at $\gamma_z=500$, controlled by the density of $Z$ flags. (c) Unimodal distribution of $Z$ flag density changes into (d) bimodal distribution at $\gamma_x=18.25$.
}
\end{figure}

\subsection{First order transitions} 
\label{sec:D4transition}

Finally, we study the nature of the transitions separating these different steady-state phases, which we find to be first order. In the previous Section, we established that the active to absorbing state transition of the flags coincides precisely with the transition of the stabilizer system. As it is numerically less expensive to simulate the reduced dynamics of flags, here we exploit the aforementioned fact and present numerical results based on simulations of only the flag densities, allowing access to higher system sizes.

We emphasize that the flags are classical degrees of freedom and undergo a \textit{classical} non-equilibrium absorbing state transition~\cite{hinrichsenNonequilibriumCriticalPhenomena2000,henkelNonEquilibriumPhaseTransitions2008,odor2004review}. The density of active sites and the survival probability of the initial activity serve as order parameters for such transitions. The active phase corresponds to a non-vanishing probability of persistent activity out to times exponentially long in the system size, whereas the system becomes inactive after a finite time in the absorbing phase.  
The location of this transition can be determined from the time-scale after which activity disappears. In the present case, this can be determined from the time it takes for the density of active (i.e., unflagged) sites, given by $1-n^f_X$ or $1-n^f_Z$, to fall to small values -- this changes from exponential in system-size in the active phase, to a constant value independent of system-size in the absorbing phase. 

As in the equilibrium case, these transitions may exhibit a first-order or second-order character. However, unlike in the equilibrium case, there is no notion of free energy for such absorbing state transitions~\cite{odor2004review}. Nonetheless, we can still make an analogy to the equilibrium scenario by monitoring the behavior of order parameters. For instance, when the system undergoes a continuous transition (e.g., in the case of directed percolation), the spatial and temporal correlation lengths diverge at the critical point~\cite{hinrichsenNonequilibriumCriticalPhenomena2000}. This results in the power-law decay of density of active sites at the critical point, which can be analyzed using standard scaling techniques analogous to equilibrium critical phenomena~\cite{henkelNonEquilibriumPhaseTransitions2008}. As we show in the following though, near the transition point of the models considered here, a significant fraction of Monte Carlo trajectories enter the absorbing state while the remaining trajectories maintain an active configuration. Such bistable behavior is instead more analogous to first-order equilibrium transitions~\cite{binder1987review}. Analysis of the discontinuous transitions to the absorbing state using the finite-size scaling methods generally involves modifications to the dynamics itself~\cite{oliveira2015finitesize}, which we do not pursue here. Instead, we observe that the system exhibits a bistability of both the absorbing and the active (with much lower flag density) configurations -- also observed in various discontinuous absorbing phase transitions~\cite{lubeck2006tricritical,grassberger2006tricritical,henkelNonEquilibriumPhaseTransitions2008}, and qualitatively resembling equilibrium first-order transitions.

To study the nature of the transition, we begin by locating it precisely. Here, we will determine the location of the transition along the $\gamma_z=500$ cut in the phase diagram shown in Fig.\ref{fig:2dphase}(b). For this, we exploit the fact that the time $\tau$ to reach a large flag density (in this case, $0.65$) grows exponentially in system size in the active phase, while in the absorbing phase it reaches an $L$-independent value for modest system sizes. We can thus use the change in the concavity of the $\tau$ vs $L$ curve on a log-log scale to pinpoint the transition from the absorbing to the partially X-active phase. We note that the dynamics of $X$ flags on the honeycomb lattice of a fixed color is independent of $X$ flags on the remaining lattices and all of the $Z$ flags. Hence, we locate this transition by only simulating the $X$ flags on a single honeycomb lattice, which allows access to higher system sizes.  In Fig.~\ref{fig:first-order-tau}(a), we see that for $\eta=1,\gamma_z=500$, this happens at $\gamma_x=6.45\pm0.03$ independent of the value of $\gamma_z$. Similarly, the transition from the partially active into the fully active state is observed at $\eta=1,\gamma_z=500,\gamma_x=18.25\pm0.25$, as shown in Fig.~\ref{fig:first-order-tau}(b). 

Having located the two transitions precisely, we now study the nature of the transition by monitoring the density of the $X$ and $Z$ flags as we cross each transition point. The probability distribution of the density of $X$ flags at the first transition point ($\gamma_x=6.45$) is shown in Fig.~\ref{fig:x-flag-hist}(b). The snapshot is taken at a time $t$ such that $\langle n^f_X(t)\rangle_{avg}=0.6$. We observe a bimodal structure: in a significant fraction of samples, the system has maintained a density of flags that is much smaller than $1$, leading to a peak at a low flag density $n^f_X \approx 0.2$; in the remaining samples, the system is already in the absorbing state, leading to a $\delta$-function peak at $n^f_X = 1$. These two peaks are well separated in the space of flag densities, indicating that the probability of being in a state with an intermediate flag density is vanishingly small. As discussed above, such bistable behavior is a hallmark of first-order transitions, where the order parameter changes discontinuously by shifting the relative weights of two distinct configurations~\cite{binder1987review,oliveira2015finitesize}.  

The bistability is observed for a finite window of time, after which most of the trajectories are trapped in the absorbing state due to finite size effects. The time dependence of the average flag density shown in Fig.~\ref{fig:first-order-tau} suggests that this time-scale grows polynomially with $L$.  To see that this is indeed the case, in the inset of Fig.~\ref{fig:x-flag-hist}(b) we show the density of $X$ flags computed by averaging only over those realizations that have not entered the absorbing state at a given time. This corresponds to the density of flags contained in the realizations near the secondary peak in Fig.~\ref{fig:x-flag-hist}(b). We observe that while the average density decays as a function of time, the conditioned density remains stable. The density among the surviving trajectories remains at a fixed value for a long time, indicating the robustness of the peak at low flag density $\langle n^f_X\rangle=n_0$. This demonstrates that (given the relatively narrow peak-widths) we can approximate the total density as a weighted average 
\eq{}{n^f_X(t) = P_\text{absorb}(t) + n_0(1-P_\text{absorb}(t)) \, ,}
where $P_\text{absorb}(t)$ is the fraction of trajectories that are in the fully occupied absorbing state. Thus, the time-scale over which bistable behavior is observed (i.e. $P_\text{absorb}<1$) for a given system size is the same as $\tau$ estimated from the full densities in Fig.~\ref{fig:first-order-tau}-- that is, the time window grows as $poly(L)$ near the transition. In contrast, for $\gamma_x$ values deep in the active phase, the time to reach an absorbing configuration grows exponentially with system size.

We note that this bimodality is sharply distinct from the time evolution seen for systems in the absorbing phase. In this case, at any given time, the probability distribution is characterized by a single peak, which moves to the right (i.e., to higher $n^f_X$) with increasing time, until all the weight is concentrated on the absorbing state. Fig.~\ref{fig:x-flag-hist}(a) shows this for $\gamma_x=4$ (in the absorbing phase) evaluated at the time $t'$ ($t' \neq t$), for which the flag density $\langle n^f_X(t')\rangle_{avg}=0.6$ is equal to that shown in Fig.~\ref{fig:x-flag-hist}(b). 

Figs.~\ref{fig:x-flag-hist}(c-d) show similar data for the densities of $Z$ flags near the transition from the partially active phase (which is an absorbing state for the $Z$ flags) into the fully active phase. Again, bistable behavior is observed near the transition (panel (d)) while in the $Z$-absorbing phase, a single peak is observed about the mean flag density, which moves to the right as time increases until all the weight is in the $Z$ absorbing state $\langle n^f_Z  \rangle= 1$.


\begin{figure}
\includegraphics[width=\columnwidth]{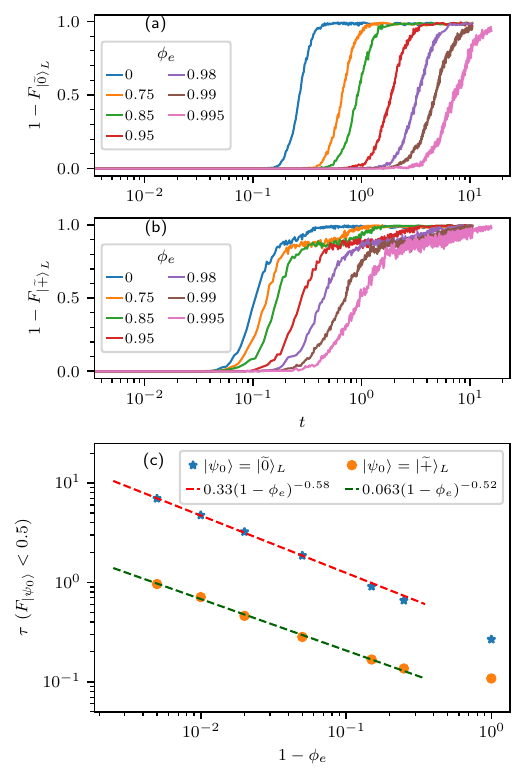}

\caption{\label{fig:d4-unheralded} $D_4$ model in the presence of unheralded noise: the logical error probability $1-F$ of system affected by a combination of heralded (rate $\phi_e$) and unheralded (rate $(1-\phi_e)$) noise, along with the correction channel, is plotted as a function of time for varying values of $\phi_e$. Here, $F_{\psi_0}$ is the fidelity of the initial logical state $|\psi_0\rangle$ with the decoded state. The data is shown for $L=18$ sized system, averaged over $250-500$ Monte Carlo realizations where all flags are initialized in the $n^f_{X_j}=n^f_{Z_j}=0$ state. The logical error probabilities (computed using the MWPM decoder) are shown for qubits initialized in (a) $|\widetilde{0}\rangle_L$ and (b) $|\widetilde{+}\rangle_L$ logical states. (c) The time $\tau$ required for the fidelity to decrease below $1/2$ is plotted as a function of the unheralded noise fraction. The dashed lines are obtained by fitting the data points (except $\phi_e=0$) to a power-law $\tau=a(1-\phi_e)^b$.}
\end{figure}

\subsection{Imperfect heralding} 
\label{sec:d4-unheralded}

Thus far, we have only considered the situation where all error processes are detectable and can be perfectly converted into erasures by raising the flags. Under such idealized conditions, we demonstrated that perfect heralding of errors provides local information that is sufficient for correcting them and stabilizing a topologically ordered steady-state. In practice, however, some of the errors are not converted into erasures due to experimental imperfections \cite{maHighfidelityGatesMidcircuit2023}.
Moreover, qubits are also affected by error mechanisms that do not take them out of the computational subspace; these events can't be detected by measurement without disturbing the state of qubit \cite{wu2022,kubica2023}. Here, we investigate the impact of introducing a low rate of unheralded errors, as will realistically be the case for even the most promising erasure-qubit platforms.

The specific nature of these unheralded noise processes depends on the details of the qubit platform. Here, we phenomenologically model them as an uncorrelated depolarizing channel acting on qubits without raising any of the erasure flags. Let $\phi_e$ be the fraction of errors that are heralded by flags and converted into erasures (i.e., $\phi_e = 1$ corresponds to the perfect heralding situation being considered so far). Then, the Lindblad superoperator for unheralded error is given by
\eq{}{\mathcal{L}_{\eta_p}(\rho)=\frac{\eta(1-\phi_e)}{3}\sum_{j}(X_j\rho X_j + Y_j\rho Y_j+Z_j\rho Z_j-3\rho) \, ,}
where $j$ runs over all qubits in the system. The heralded fraction of the noise is represented by the erasure noise model in Eq.~\eqref{eq:tc-noise}, where we replace $\eta\rightarrow\eta\phi_e$. The time evolution of the state under the combined heralded and unheralded noise, along with correction operations, is generated by $\mathcal{L}_{\eta_e}+\mathcal{L}_{\eta_p}+\mathcal{L}_{C}$. The unheralded errors pair-create stabilizer defects that are not connected by strings of flags. Our correction protocol does not act on such defects if there are no flags in their vicinity. Hence, once such defects are created their evolution is effectively uncorrelated from that of their partner. In a long time limit, for any finite value of unheralded error rate, this results in a finite density of defects that are well-separated from their partners. We expect that upon pairing these defects using MWPM decoder, we will not get back to the initial state with certainty, meaning that our steady state has lost its topological properties. In Fig.~\ref{fig:d4-unheralded}(a)-(b), we demonstrate this by evaluating the fidelity of the state obtained by decoding the time-evolved state using MWPM decoder with respect to two different initial states in the $D_4$ ground-state space. The fidelity in the steady-state vanishes for all values of erasure fraction $\phi_e<1$.

While the steady-state quantum memory is unstable against unheralded errors, however, when most of the errors are heralded, the correction protocol nevertheless results in a significant enhancement of the time-scale up to which the encoded information can be recovered.   Specifically, the time required for the fidelity to decay to a given value increases with increasing value of the heralding fraction $\phi_e$ as $\tau \sim (1- \phi_e)^{-\alpha}$, where numerically we observe that $\alpha \approx 0.5$. 

\begin{figure}
\subfloat{%
  \includegraphics[width=\columnwidth]{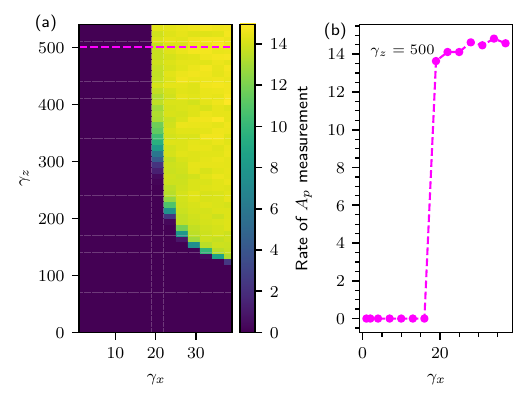}
}
\caption{\label{fig:d4-a-checks} (a) The rate at which plaquette stabilizers $A_p$ are measured in the steady-state is shown for perfectly heralded dynamics in the $\gamma_x-\gamma_z$ parameter space of the $D_4$ model. We compute this rate in the Monte Carlo (MC) simulation of flags by counting the average number of $A_p$ measurements per plaquette proposed by the correction jump operators in a single MC sweep. The data is shown for an $L=48$ sized system at a time $t_f=20$ and averaged over $100$ MC realizations. The measurement rate goes to zero in the absorbing and partially active phase (purple region) since all $Z$ flags are raised in this regime. In the active phase (yellow region), the plaquette measurements are applied at a finite rate by the correction channel. (b) The measurement rate is shown for a fixed value of $Z$ correction rate $\gamma_z=500$ as a function of $X$ correction rate (horizontal cut along dashed line in (a)). The measurement rate jumps from zero to a finite value near the transition where the steady-state density of $Z$ flags changes discontinuously.}
\end{figure}
\section{Conclusions}
\label{sec:cncls}

In this paper, we have presented a fully local dynamical protocol that, under the assumption of perfectly heralded noise, can stabilize steady-state phases with Abelian ($\mbZ_2$) or non-Abelian ($D_4$) topological order in two dimensions. Building on our previous work in 1d~\cite{chirame2024spt}, here we have demonstrated that the information about error locations provided by erasure flags can be exploited to effectively confine the errors, leading to a non-trivially ordered active dynamical phase for noise rates below a certain threshold. After illustrating our correction protocol for the Toric code, we have used three different metrics to show that the active phase in the $D_4$ model constitutes a legitimate dynamical phase of matter. First, we have verified that this phase effectively preserves quantum information, in the sense that the logical information can be faithfully recovered by applying the MWPM decoder. Second, we have provided compelling numerical evidence that this active phase lies in the same mixed-state phase as the pure $D_4$ ground state: specifically, we show that a local quantum channel can take a state in this active phase and return it to a pure $D_4$ ground state in a time that scales at most logarithmically in the system size. Finally, we have shown that a first-order phase transition separates the dynamics of flag and defect distributions in this active phase from neighboring dynamical regimes. Besides the fully decohered absorbing phase, we have also established the existence of an intermediate steady-state phase, wherein classical (but not quantum) information can be preserved with high fidelity.

Several aspects of our results are worth emphasizing.  First, we have tested our model in the presence of \textit{unbiased} heralded noise, even though heralded errors most often occur in systems where the noise is highly biased. As such, unbiased noise presents the most challenging obstacle for our non-Abelian decoder, since there is a high probability that any given application of the noise channel creates new non-Abelian defects. In a scenario where the noise is biased, such that $Z$-type errors occur more frequently than $X$-type errors, the majority of the errors created will be Abelian plaquette defects, which require considerably less overhead to correct. Under these conditions, the $Z$ correction rate required to maintain a fully active $D_4$ phase would be much closer to the values seen for the Toric code than those observed for the $D_4$ model with unbiased noise.   

Second, our local decoder exploits the fact that the fusion rules of the $D_4$ topological order are acyclic, meaning that two non-Abelian anyons of the same type fuse only to Abelian anyons. This property allows us to remove non-Abelian defects by applying a sequence of single-site $X$ operators. While advantageous for a general non-Abelian decoder (such as the MWPM protocol used in Sec.~\ref{sec:D4}), this feature is essential for our local decoder, since the string operators that pair-annihilate non-Abelian anyons {\it without} generating further excitations along their length are high-depth unitary circuits that cannot be decomposed into a sequence of purely local operations applied independent of the correction history. In replacing these high-depth unitary circuits with strings of onsite $X$ operators, the main price that we pay is the creation of large numbers of Abelian anyons along the strings' length; these however can be corrected locally at sufficiently high $Z$-correction rates. In principle, we expect that our correction protocol can be generalized to other acyclic anyon theories e.g., those described by $\mathcal{D}(G)$, where $G$ is some class-$n$ nilpotency group (where $n$ is finite). It would be interesting to investigate the conditions under which these generalizations also admit fully active phases that remain topologically ordered, leading to self-correcting quantum memories when the noise is perfectly heralded. It remains an open question whether our protocol generalizes to the case when $G$ is solvable but not nilpotent (e.g., $S_3$), where state preparation remains efficient under measurement and feedforward~\cite{nathierarchy}; this question is particularly interesting since anyon theories based on solvable but not nilpotent groups $G$ can be used for universal topological quantum computation~\cite{mochon2004,galindo2018}.

Third, in realistic experiments (amongst other imperfections), the measurement-plus-feedback process will itself be error prone, and it is natural to wonder how our protocol fares in this context. As exhibited by the robustness of the fully active phase to $X$ correction processes, which introduce additional heralded errors, these new types of errors are not expected to qualitatively alter the phase diagram, provided that the errors remain perfectly heralded. However, this is not a realistic expectation for current experiments; instead, we conclude that measurement failures are likely to introduce unheralded errors, which will destroy the steady-state order and lead instead to a finite lifetime for the topological memory. Nonetheless, as we show in Fig.~\ref{fig:d4-unheralded}, our correction protocol enhances the lifetime of the encoded quantum information when the rate of unheralded errors remains low. 

Our results also provide insight into steady-state phases for models in a TO phase away from their fixed points (such as the Toric code or $D_4$ model discussed here). The ground state of any point in the TO phase is connected to the respective fixed-point ground state by finite-depth quasi-local unitary operators~\cite{hastings2005quasiAdiabatic,chen2010localUnitary,bravyi2010}. This, in turn, means that the code-space of such a general point in the TO phase will be defined by quasi-local quasi-stabilizers. As a result, single-qubit errors will create more quasi-stabilizer violations compared to the fixed-point state. However, if these errors are perfectly heralded, then the resulting quasi-stabilizer violations will still be connected by strings of flags on some coarse-grained level. We expect that an appropriately modified version of our protocol can successfully target a stable active phase even in this case.

From a practical point of view, we also briefly commented on the feasibility of achieving the correction rates necessary to stabilize the TO phase on current hardware. It turns out that the quantum memory phase requires a relatively large rate $\gamma/\eta\approx 500$ of correction processes involving entangling gates. In practice, however, the stabilizer measurements in our protocol have to be performed only when the classical erasure flags are in an appropriate correctable configuration. Thus, the actual rate at which the entangling gates are implemented, particularly during $Z$-correction protocol, is expected to be lower than the bare correction rate $\gamma_z$. In Fig.~\ref{fig:d4-a-checks} we show the rate at which the measurement and feedback operations relevant for $Z$ correction are performed (i.e., measurement of $A_p$) at various points in the phase diagram, revealing that the actual gate implementation rates are much lower than the bare correction rates throughout the active phase. 

Finally, a technical contribution of this work is an extension of the conventional stabilizer tableau to non-Pauli quasi-stabilizers, which permits efficient simulation of open quantum dynamics for the $D_4$ model. A necessary ingredient is that the quasi-stabilizers commute when restricted to the ground-state subspace (i.e., the ground-space is frustration-free). Further, the adjoint actions of the noise channels considered here map local operators to local operators. In the future, it would be interesting to understand more generally which properties of the noise and recovery jump operators are necessary in order for our generalized stabilizer tableau formalism to apply.


\begin{acknowledgements}

A.P. is grateful to Ramanjit Sohal and Nat Tantivasadakarn for helpful conversations. S.G. thanks Curt von Keyserlingk, Tibor Rakovszky, and Jeff Thompson for helpful discussions. The authors acknowledge helpful conversations with David Huse. F.J.B. and S.C. are grateful for the support of NSF DMR-2313858. S.G. acknowledges funding from an Institute for Robust Quantum Simulation (RQS) seed grant. This material is based upon work supported by the Sivian Fund and the Paul Dirac Fund at the Institute for Advanced Study and the U.S. Department of Energy, Office of Science, Office of High Energy Physics under Award Number DE-SC0009988 (A.P.). The authors acknowledge the computational resources provided by the Minnesota Supercomputing Institute (MSI) at the University of Minnesota.
    
\end{acknowledgements}

\section*{Data availability} All simulation codes and generated datasets are available online at \cite{chirame2025simulationDatad4}.

\appendix


\begin{figure}
  \includegraphics[width=\columnwidth]{{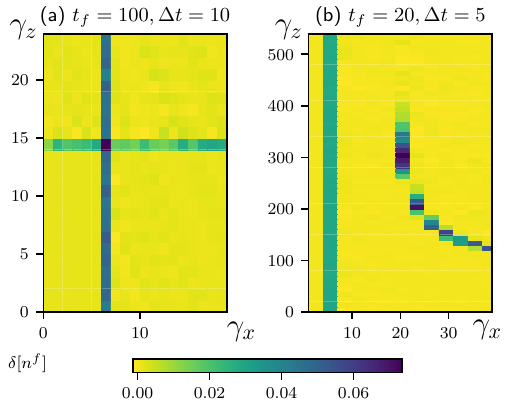}}%
\caption{\label{fig:additional-comparison} The change in the total flag density $\delta[n^f]$ (see Eq.~\eqref{eq:vardens}) over a time interval $\Delta t$ is shown for (a) the Toric code (b) the $D_4$ model in the $\gamma_x-\gamma_z$ parameter space at some late time $t_f$. In both plots, the change in the value of density after time $\Delta t$ is small for parameters away from the phase boundaries. This suggests that the late time values evaluated at $t_f$ (see corresponding phase diagrams for flag densities in Fig.~\ref{fig:2dphase}) are a good estimate of the steady-state densities. Near the transition, determination of the steady-state values requires more careful analysis of larger system sizes (presented for one specific value $\gamma_z=500$ in Fig.~\ref{fig:first-order-tau} to identify the location of transition).}
\end{figure}

\section{Details of numerical Monte Carlo simulations}
\label{app:numerics}

In this Appendix, we provide the details of the Monte Carlo algorithm used to simulate the Lindblad dynamics. The system is time evolved using random sequential updates (see eg., Ref.~\cite{hinrichsenNonequilibriumCriticalPhenomena2000}) on the state of the flags and the stabilizer tableau. Here, we present a rough outline of the procedure for the local noise+correction dynamics of the $D_4$ model; the Toric code is simulated in an analogous manner.

The flags are initialized in the state $n^f_X=n^f_Z=0$ for all sites and the stabilizer tableau is initialized in one of the logical states of the $D_4$ stabilizers (see Sec.~\ref{sec:d4quasi}). Recall that $\phi_e$ corresponds to the fraction of heralded errors (i.e., $\phi_e = 1$ refers to perfect heralding) and that $\gamma_x,\gamma_z$ parameterize the rate of $X$ and $Z$ correction, respectively. Then, the state is evolved up to time $t_f$ as follows:
\begin{algorithmic}
\While {$t\leq t_f$}
\For{$i=1,2,\ldots 3L^2$}
    \State sample a random number $1\leq j\leq 3L^2$ from the uniform distribution to choose the location of the qubit on which $\mathcal{L}_\eta$ acts 
    \State sample a binary random number ($v_j$ and $p_j$) to choose one of the two vertices/plaquettes connected to edge $j$ where $\mathcal{L}_C$ will act
    \State sample another random number to choose the update rule for the state (with $c_0:=1/(\eta+2\gamma_x/3+\gamma_z/3)$ here):
    \State with probability $c_0\eta\phi_e/4$: Apply $X_j$ and set $n^f_{X_j}=n^f_{Z_j}=1$
    \State with probability $c_0\eta\phi_e/4$: apply $Z_j$ and set $n^f_{X_j}=n^f_{Z_j}=1$
    \State with probability $c_0\eta\phi_e/4$: apply $Z_j\times X_j$ and set $n^f_{X_j}=n^f_{Z_j}=1$
    \State with probability $c_0\eta\phi_e/4$: set $n^f_{X_j}=n^f_{Z_j}=1$
    \State with probability $c_0\eta(1-\phi_e)/3$: apply $X_j$ 
    \State with probability $c_0\eta(1-\phi_e)/3$: apply $Z_j$ 
    \State with probability $c_0\eta(1-\phi_e)/3$: apply $Z_j\times X_j$ 
    \State with probability $2c_0\gamma_x/3$: apply the appropriate jump operator from the $X$ correction move-set that is consistent with the flag structure at vertex $v_j$ 
    \State with probability $\ c_0\gamma_z/3$: apply the appropriate jump operator from the $Z$ correction move-set that is consistent with the flag structure at plaquette $p_j$ 
\EndFor
\State $t\rightarrow t+\delta t:=t+\frac{1}{\eta+2\gamma_x/3+\gamma_z/3}$
\State Record the values of the observables in the state at the end of the Monte Carlo sweep
\EndWhile
\end{algorithmic}

Throughout this work, we set the rate of noise $\eta = 1$. Note that the number of Monte Carlo sweeps required to simulate the system up to a fixed time $t_f$ increases for increasing values of correction rates. We present additional data in Fig.~\ref{fig:additional-comparison} to show that the final time $t_f$ chosen to compute the phase diagram in Fig.~\ref{fig:2dphase} provides a good estimate of the densities in the genuine long-time steady-state. In particular, we verify that the variation of the total density over some time interval $\Delta t$
\be
\label{eq:vardens}
\delta[n^f]:=n^f(t=t_f)-n^f(t=t_f-\Delta t) \, ,
\ee
where $n^f=\frac{1}{2}{(n^f_X+n^f_Z)}$, remains small at late times $t_f$ (away from the phase boundaries).


\section{Review of $D_4$ Topological Order}
\label{app:D4review}

The topological order displayed by the Hamiltonian Eq.~\eqref{Eq:D4Ham} is encoded in the twisted quantum double $\mathcal{D}^\alpha(\mbZ_2^3)$, which is equivalent to that of the quantum double of $\mathcal{D}(D_4)$. Here, we review first the anyon content of $\mathcal{D}(D_4)$ and then provide a mapping to the anyon content of $\mathcal{D}^\alpha(\mbZ_2^3)$, which is the convention used throughout the main text. 

The group $D_4 \cong (\mbZ_2 \times \mbZ_2) \rtimes \mbZ_2 $ corresponds to the symmetries of a square and is specified abstractly by elements $r,s$:
\begin{align}
    D_4 = \left\langle r,s \, \large| \, r^2=s^4=rsrs=1
    \right\rangle 
    \, ,
\end{align}
with center $\langle s^2 \rangle$. There are five conjugacy classes $\{1\}$,  $\{s^2\}$,  $\{s,s^3\}$,  $\{r,r s^2\}$,  $\{rs,rs^3\}$, and hence also five irreducible representations (irreps): four of dimension one and one of dimension two. We label the one dimensional irreps by $ij$ ($i,j \in \mbZ_2$). Besides the trivial irrep $00$ and the faithful two dimensional irrep $\sigma$, there are three sign irreps:
\begin{itemize}
    \item $10$ is represented by $r = sr = s^2 r = s^3r = -1$,
    \item $01$ is represented by $s = s^3 = sr = s^3r = -1$,
    \item $11$ is represented by $s = s^3 = r = s^2 r = -1$.
\end{itemize}
The character table is given by 
\begin{align}
\begin{tabular}{ l r r r r r }
 & $1$  & $s^2$ & $s$ & $r$ & $rs$  \\
 \hline
$\chi_{00}$ & 1  & 1 & 1 & 1 & 1 \\
$\chi_{01}$ & 1 & 1 & $- 1$ & 1 & $-1$  \\
$\chi_{10}$ & 1 & 1 & 1 & $-1$  & $-1$  \\
 $\chi_{11}$ & 1 & 1  & $-1$  & $- 1$ & 1 \\
$\chi_\sigma$ & 2 & $-2$ & 0 & 0 & 0
\end{tabular}
\label{chartbl}
\end{align}
Notice that the first four irreps obey the obvious $\mathbb{Z}_2 \times \mathbb{Z}_2$ fusion rules, while 
\begin{align}
   {ij} \otimes \sigma = \sigma \, ,
    &&
    \sigma \otimes \sigma = \sum_{i,j} ij \label{eq:sigma fusion}
    \, ,
\end{align}
obey a $\mathbb{Z}_2$ grading.

The anyon content of $D_4$ TO is specified as follows: each point-like anyon can be labelled by the ordered pair $([g],\pi_g)$, where $[g]$ corresponds to a conjugacy class and $\pi_g$ denotes an irrep of its centralizer~\cite{dijkgraaf}. Thus, the four pure flux excitations are locally labelled by the non-trivial conjugacy classes, while the four pure charge excitations are labelled by the non-trivial elements of Rep$(D_4)$. The remaining thirteen non-trivial excitations are dyons, leading to a total of twenty-two anyons (including the trivial anyon) in $D_4$ TO. The braiding phase between the pure charges and fluxes is given by the phase of the corresponding entry in the character table above: for instance, the wave function acquires a minus sign upon braiding the $11$ charge around the $[s]$ flux or when the $\sigma$ charge is braided with the $[s^2]$ flux.

The excitations of the Hamiltonian Eq.~\eqref{Eq:D4Ham} considered in the main text can be labelled similarly to those of three copies of the 2d Toric code; these are generated by $e_c$ and $m_c$, where the color index $c \in \{r,g,b\}$ indexes the different copies. Recall that the anyon content of the 2d Toric code is generated by the bosons $e$ and $m$, which satisfy $\mbZ_2 \times \mbZ_2$ fusion rules. $e$ and $m$ have non-trivial mutual statistics since braiding one around the other generates an overall minus sign, and their fusion product $\psi = e \times m$ is an Abelian fermion. As discussed in Ref.~\cite{iqbal2024nonabelian}, the 22 anyons of the twisted quantum double $\mathcal{D}^\alpha(\mbZ_2^3)$ are:
\begin{enumerate}
    \item There are three fundamental bosons $e_c  (c = r,g,b)$, which are associated with the three colors of plaquette stabilizer defects in our model. Combinations of these also yield the three bi-colored bosons $e_{rg} = e_r \times e_g$, etc., and one tri-colored boson $e_{rgb}= e_r \times e_g \times e_b$. The eighth Abelian boson is the trivial particle ($e_c^2 = 1 \, \forall \, c$).
    \item Non-Abelian bosons:
        \begin{itemize}
            \item Our three colors of vertex defects correspond to the three fundamental non-Abelian bosons $m_r,m_g,m_b$. Here, $m_c$ and $e_c$ are mutual semions, while $m_c$ braids trivially with $e_{c'}$ ($c\neq c'$).
            \item Combinations of adjacent vertex defects on different colored lattices yield an additional three non-Abelian bosons, $m_{rg},m_{gb},m_{rb}$: $m_{c c'}$ (with $ c\neq c'$) is a mutual semion with respect to $e_c$ and $e_{c'}$, but has trivial mutual statistics with $e_{c''}$ ($c \neq c' \neq c''$).
        \end{itemize}
    \item Non-Abelian fermions:
        \begin{itemize}
            \item $\psi_c = m_c \times e_c$.
            \item $\psi_{cc'} = m_{cc'} \times e_c = m_{cc'} \times e_{c'}$ (for $c \neq c'$).
        \end{itemize}
    \item A non-Abelian semion $s_{rgb}$ which braids non-trivially with $e_c \, \forall \, c \in \{r,g,b\}$.
    \item A non-Abelian anti-semion $\bar{s}_{rgb} = s_{rgb} \times e_{rgb}$.
\end{enumerate}
Each of the non-Abelian anyons in the theory has quantum dimension $d_\alpha = 2$. Ref.~\cite{iqbal2024nonabelian} provided an explicit mapping between the anyon theory of the quantum double $\mathcal{D}(D_4)$ and the anyon content of twisted $\mbZ_2$ gauge theory; we reproduce it here in Table~\ref{tab:mapping} for completeness. 

\begin{table}[t]
    \centering
        \begin{tabular}{|c|c|c||c||c|}
        \hline
 \multicolumn{3}{|c||}{$\mathcal D(D_4)$} & \multirow{2}{*}{$\mathcal D^\alpha(\mbZ_2^3)$} & \multirow{2}{*}{Dim.}\\
 \cline{1-3}
Conj. class & Centralizer & Irrep & & \\
    \hline
          $1$ &$D_4$ & $00$  & 1      & 1 \\
      $1$ &$D_4$ &  $10$  & $e_{rg}$      & 1  \\
        $1$ &$D_4$ &  $01$ & $e_{r}$      & 1 \\
  $1$ &$D_4$ & $11$  & $e_{g}$       & 1 \\
        $1$ &$D_4$ & $\sigma$  & $m_b$       & 2  \\
\hline
          $[{s^2}]$ &$D_4$ & $00$  & $e_{rgb}$   & 1 \\
$[{s^2}]$ &$D_4$ &$10$  & $e_{b}$   & 1 \\
     $[{s^2}$] &$D_4$ &  $01$ &$e_{gb}$    & 1 \\
  $[{s^2}]$ &$D_4$ &$11$ &$e_{rb}$  & 1 \\
      $[{s^2}]$ &$D_4$ &$\sigma$  &$\psi_b$ & 2 \\
\hline
        $[{s}]$ &$\mbZ_4$ & $1$ & $m_{rg}$  & 2 \\
             $[{s}]$ &$\mbZ_4$ & ${\mathbf \omega}$ & $ s_{rgb}$  & 2 \\
   $[{s}]$ &$\mbZ_4$ & ${\mathbf \omega^2}$ & $\psi_{rg}$  &2 \\
   $[{s}]$ &$\mbZ_4$ & $\overline{{\mathbf \omega}}$ & $\bar s_{rgb}$ & 2 \\
\hline
               $[{r}]$ &$\mbZ_2^2$ & $1$ &$m_{gb}$ & 2 \\
              $[{r}]$ &$\mbZ_2^2$ &$(-1,1)$&$m_{g}$ & 2 \\
     $[{r}]$ &$\mbZ_2^2$ & $(1,-1)$ &$\psi_{g}$& 2 \\
       $[{r}]$ &$\mbZ_2^2$ & $(-1,-1)$ &$\psi_{gb}$& 2  \\
\hline
         $[{sr}]$ &$\mbZ_2^2$ &$1$ &$m_{rb}$ & 2 \\
     $[{sr}]$ &$\mbZ_2^2$ & $(-1,1)$ &$m_{r}$& 2  \\
    $[{sr}]$ &$\mbZ_2^2$ &$(1,-1)$&$\psi_{r}$& 2  \\
 $[{sr}]$ &$\mbZ_2^2$ & $(-1,-1)$& $\psi_{rb}$& 2 \\
    \hline
    \end{tabular}
    \caption{Mapping between anyons in $D_4$ TO and anyon labels used in the main text, which correspond to anyons in twisted $\mbZ_2^3$ gauge theory.}
        \label{tab:mapping}
    \end{table}


\section{Simulation using Non-Pauli stabilizer tableau}
\label{StabApp}
In this Appendix, we provide details for simulating the Lindbladian dynamics of the $D_4$ model defined in terms of non-Pauli stabilizers (see Eq.~\eqref{Eq:D4Ham}). For an $N$ qubit system, the time evolution of standard Pauli stabilizers (and consequently, the states that they stabilize) can be simulated by tracking the time evolution of the $O(N)$ stabilizers, each of which is a local product of Pauli operators. Since dynamics generated by Clifford gates maps Pauli stabilizers to Pauli stabilizers, any Clifford-based dynamics can be simulated efficiently by simply updating the stabilizers at each time step~\cite{gottesman1998heisenberg}. The ground-space of the $D_4$ model is instead stabilized by the non-Pauli stabilizers (or ``quasi-stabilizers") defined in Eqs.~\eqref{eq:d4-b-stab} and~\eqref{eq:d4-a-stab}: each of these is a tensor-product of $1$ and $2$ qubit operators that are localized around a single vertex or plaquette of the honeycomb lattice. In an abuse of terminology, we will refer to these operators simply as stabilizers, but we stress that these are not mutually commuting elements of the $N$-qubit Pauli group. We now provide a generalized protocol that shows how the evolution of these stabilizers under the Pauli operations present in our dynamical protocol can be tracked similarly to genuine stabilizers, allowing us to efficiently simulate the noise and correction Lindbladian dynamics of the $D_4$ model.  

We begin by choosing a basis for the independent stabilizers of the initial $D_4$ state. Any ground state of the $D_4$ Hamiltonian Eq.~\eqref{Eq:D4Ham} is fully fixed by a list of plaquette and vertex stabilizer generators, together with the non-local logical operators that stabilize it. Each vertex of the honeycomb lattice has a vertex stabilizer $B_v$ (see Eq.~\eqref{eq:d4-b-stab}). While there are $2L^2/3$ operators on each sublattice of a system with $L \times L$ lattice sites, for each color $c$, these obey a global constraint $\prod_{v_c} B_{v_c}=\mathbb{I}$. Hence, the set of independent generators can be represented by
\eq{eq:Qdefinition}{\mathcal{Q} = \{q_1 = B_{v_1},q_2 = B_{v_2},\ldots,q_{2L^2-3} = B_{v_{2L^2-3}} \} \, ,}
with the values of the remaining $3$ vertex stabilizers fully determined by this linearly independent set.

Now, while the plaquette stabilizers do not satisfy a similar constraint on the full Hilbert space i.e., $\prod_{p_c}A_{p_c}\neq \mathbb{I}$, they are however constrained within the vertex-defect free subspace (i.e. $B_{v_c}\equiv 1$) via the relation:
\eq{Eq:PlaquetteRedundancy}{\prod_{p_c}A_{p_c} = (-1)^{\frac{(1-\mathcal{Z}^{c'}_H)}{2}\frac{(1-\mathcal{Z}^{c''}_V)}{2}}\times (-1)^{\frac{(1-\mathcal{Z}^{c'}_V)}{2}\frac{(1-\mathcal{Z}^{c''}_H)}{2}} \, ,
} 
where the product is taken over all plaquette operators of the same color $c$ and $\mathcal{Z}_{H/V}^{c'/c''}$ represents the logical-$Z$ operators of the remaining colors $c'\neq c''\neq c$ in the horizontal (vertical) direction. Since the ground-space is by definition defect-free, the set of independent plaquette generators relevant to our initial state can be represented as:
\eq{eq:Gdefinition}{\mathcal{G} = \{g_1 = A_1,g_2 = A_2 ,\ldots,g_{L^2-3} = A_{L^2-3} \} \, ,}
with the values of the remaining plaquette stabilizers, denoted $A_{p_c^*}$ (one for each $c$), fixed by Eq.~(\ref{Eq:PlaquetteRedundancy}).

While the above relation Eq.~\eqref{eq:Gdefinition} holds true for a $D_4$ ground-state, it will not be valid in the time-evolved state in general since this can have vertex stabilizer violations, generated by the noise. Instead, the relation will be modified as a result of the stabilizer updates. In principle, we can infer the modified relation at each time-step from the set of instantaneous independent stabilizers. However, implementing this in practice requires decomposing each plaquette operator $A_p$ in terms of its constituent $CZ$ gates to determine their global pattern (as was done to derive Eq.~\eqref{Eq:PlaquetteRedundancy}): this defeats the goal of treating operator strings as tensor products of the $A_p$ operators. To overcome this, we instead explicitly track the time evolution of three product operators $\prod_c A_{p_c}$, one for each color $c$. Each of these are length-$L^2/3$ operator strings composed of products of local $A_{p_c}$ operators, which allows their evolution to be carried out efficiently using the methods described below. 

Finally, depending on the specific logical state chosen in the simulation (see Eq.~\eqref{eq:d4-initial-state}), we choose one of the following two sets of initial logical stabilizers:  
\eq{}{
\mathfrak{L} = \begin{cases}
    \{\mathcal{Z}^r_{V},\mathcal{Z}^g_{V},\mathcal{Z}^b_{V},\mathcal{Z}^r_{H},\mathcal{Z}^g_{H},\mathcal{Z}^b_{H}\} ,\ \text{ if } |\psi_0\rangle = |\widetilde{0}\rangle_L\\
    \{\mathcal{Z}^r_{V},\mathcal{Z}^g_{V},\mathcal{Z}^b_{V},\mathcal{X}^r_{V},\mathcal{X}^g_{V},\mathcal{X}^b_{V}\} ,\ \text{ if } |\psi_0\rangle = |\widetilde{+}\rangle_L\\
\end{cases}
}
In total, the $3L^2$ operators from the combined sets $\mathcal{Q}, \mathcal{G}$, and $\mathfrak{L}$ describe a unique state as their $+1$ eigenvector~\cite{iqbal2024nonabelian}.

To efficiently simulate the measurement dynamics, we also need to define a plaquette `destabilizer' generator $d_k$ corresponding to each $g_k$~\cite{aaronson2004improved}. These operators are chosen such that they satisfy the following algebraic relations:
\eq{eq:destab-relation}{[d_j,d_k]=[d_j,g_k]=0 \text{  for }j\neq k \text{ , and  } \{d_j,g_j\}=0 }
for all $g_j\in \mathcal{G}$. Specifically, each destabilizer only anticommutes with its partner stabilizer; this allows efficient detection of the presence or absence of $g$ in operator strings, as we discuss later. The destabilizer $d_{p_c}$ for the $D_4$ initial state can be defined as 
\eq{}{d_{p_c} = \prod_{k\in l_p} Z_{k} \, ,}
where $l_p$ represents a path on the dual lattice connecting the center of plaquette $p_c$ to the designated plaquette $p^*_c$ of the same color (recall that $A_{p^*_c} \notin \mathcal{G}$). It is straightforward to verify that this choice satisfies Eq.~\eqref{eq:destab-relation} since every generator along the path $l_p$ gets acted on by two $Z$ operators, and hence they all commute with $d_{p_c}$. Only the operators $A_{p_c}$ and $A_{p^*_c}$, located at the string's  endpoints, anticommute with $d_{p_c}$; since $A_{p^*_c}\notin \mathcal{G}$, the destabilizer is uniquely associated with the plaquette stabilizer $A_{p_c}$ in the initial state.

\subsection{Evolution under $X$ and $Z$ operators}
To simulate the dynamics, we will track individual trajectories of our system over time. If $S$ stabilizes the state $|\psi (t) \rangle$ at time $t$, then the state $|\psi (t+ 1) \rangle:=U|\psi (t)\rangle$ is stabilized by the modified stabilizer operator $S':=USU^{-1}$.  This unitary evolution also affects the destabilizer operators, which also need to be updated according to $D':=UDU^{-1}$. This simultaneous transformation leaves the defining properties of destabilizers in Eq.~\eqref{eq:destab-relation} invariant.
In the case at hand, both the noise and feedback operations involve the application of Pauli $X$ and $Z$ operators. This leads to the modified $B_v$ and $A_p$ operators given in Eq.~\eqref{eq:z-stab-update}, \eqref{eq:x-Bstab-update}, and \eqref{eq:x-Astab-update}.

The logical stabilizers in $\mathfrak{L}$ can be updated using similar transformations. The logical $Z$ operator is unchanged after applying $Z$ operators, and transforms as 
\eq{}{\pm\mathcal{Z}\xrightarrow{X_k}\mp \mathcal{Z}}
if $X_k$ acts on the edge that lies on the path defining $\mathcal{Z}$. The logical $X$ operator has both $X$ and $CZ$ operators in it. The corresponding transformation rule can be obtained using Eq.~\eqref{eq:cz-relation} as
\eq{}{\pm\mathcal{X}\xrightarrow{Z_k}\mp \mathcal{X}, \text{ and} \ \pm\mathcal{X}\xrightarrow{X_j}\mp \prod_{i\in \xi(j)}Z_{i} \mathcal{X}.}
Here $Z_k (X_j)$ acts on one of the edges that have Pauli $X (Z)$ in the definition of logical $X$, and $\xi(j)$ is the set of edges coupled by $CZ$ to edge $j$ in the definition of $\mathcal{X}$.

A key takeaway from this step is that, at any time $t$, the transformed stabilizer generators $g\in\mathcal{G}$ and $q\in\mathcal{Q}$ can be written as
\eq{Eq:Stabilizersevolved}{g=\mathsf{s_g}\prod_k Z_k A_p, \text{  and   } q_v = \mathsf{s_v} B_v}
where $\mathsf{s_g,s_q}=\pm1$. These are still tensor-products of a set of $1$ and $2$-qubit operators localized near plaquettes and vertices.

\subsection{Measurement of $B_v$ and $A_p$} 

We now discuss the implementation of measurement operations. This requires determining the measurement outcome and the stabilizers of the post-measurement state.

The vertex operators $B_v$ commute with all stabilizers and hence their measurement leaves the state unchanged. The corresponding measurement outcome is readily available from the sign of elements in $\mathcal{Q}$. Similarly, if a measurement operator $A_p$ is present in the list of instantaneous generators (i.e., $\pm A_p\in\mathcal{G}$), then the outcome is given by the sign of the stabilizer and the state is not modified.

If $\pm A_p$ is not amongst the list of generators, then we need to observe its algebraic relation with the generators in order to determine the result of the measurement on the state. In particular, note that $A_{p'}$ commutes with $A_p$ up to a product of vertex operators (see eg. Eq.~\eqref{eq:d4-commute-relation}). It follows that the instantaneous stabilizer generators $g=\prod_k Z_k \prod_{p'}A_{p'}$ (that is potentially made up of multiple $A$ type operators) obey the relation:  
\eq{}{\eqsp{A_p g &= A_p(\prod_{k} Z_k\prod_{p'} A_{p'})=(\prod_k Z_k A_p') A_p ( \mathsf{s} \prod_{v \in \mathcal{V}}B_v) \\
&= g A_p \ ( \mathsf{s} \prod_{v \in \mathcal{V} }B_v)\, ,}}
where the sign $\mathsf{s}=\pm1$ and the list of vertex operators $v \in \mathcal{V}$ obtained by commuting $A_p$ with plaquette operators in $g$ can be determined using Eqs.~\eqref{eq:d4-a-stab} and \eqref{eq:d4-commute-relation}. The probability of getting a measurement outcome $\mathsf{m}=\pm1$ upon measuring $A_p$ is then given by
\eq{eq:app-pm}{\eqsp{\mathsf{p}_\mathsf{m} &:= \langle \psi|\frac{1+\mathsf{m}\ A_p}{2}|\psi\rangle = \langle \psi|g_{j}\frac{1+\mathsf{m}\ A_p}{2}g_{j}|\psi\rangle \\
&= \frac{1}{2} + \frac{\mathsf{m}}{2}  \langle \psi | A_p (\mathsf{s}  \prod_v B_v) |\psi\rangle 
}}
where we have used the fact that the state before measurement is stabilized by the operator $g_{j}\in\mathcal{G}$, with $(g_{j})^2=1$. 

\paragraph*{Case a:} If there exists at least one stabilizer $g_{j}\in \mathcal{G}$ such that $\mathsf{s}\prod_{v\in\mathcal{V}}B_v|\psi\rangle=-|\psi\rangle$, then we get 
\eq{eq:anticomm-prob}{\mathsf{p_m}=\frac{1}{2}-\frac{\mathsf{m}}{2}\langle \psi|A_p|\psi\rangle =\mathsf{p_{-m}} \, ,}
where in the last equality, we have used the fact that $\mathsf{p_{-m}} := \langle \psi | ( 1 - \mathsf{m}A_p)/2 |\psi \rangle$. Since $A_p$ is a hermitian operator that squares to $1$, $\mathsf{m}=\pm1$ are the only possible measurement outcomes which, together with Eq.~\eqref{eq:anticomm-prob}, implies that $\mathsf{p}_{+1}=\mathsf{p}_{-1}=1/2$. 
Hence, the measurement outcome in this case is equally likely to be $+1$ or $-1$, which we simulate by sampling a binary random number from the uniform distribution. Suppose $g_1,g_2,\ldots,g_{n_0}$ are the $n_0\geq 1$ stabilizers for which this holds. Then, they no longer stabilize the post-measurement state stabilized by $\mathsf{m}A_p$ since they obey $g_j(1+\mathsf{m}A_p)|\psi\rangle=(1-\mathsf{m}A_p)|\psi\rangle$. Instead, the new state of the system is stabilized by pairwise products of these generators. This leads to a modification of stabilizers as follows: we choose a particular stabilizer $g_1$ from this list and the remaining anti-commuting operators are updated as 
\eq{}{g_j\rightarrow g_1 g_j \quad\text{ for } j=2,3,\ldots n_0 \, .}
Additionally, if any of the destabilizers anticommute with the new stabilizer $\mathsf{m} A_p$, then they also have to be modified to maintain the defining relations in Eq.~\eqref{eq:destab-relation}. This is achieved by updating 
\eq{}{d_k \rightarrow g_1 d_k \quad\text{ if } \quad d_k A_p|\psi\rangle=-A_pd_k|\psi\rangle \, .}
At the end of this step, stabilizer $g_1$ is replaced based on the measurement outcome as
\eq{}{d_1 \rightarrow g_1,\qquad  g_1\rightarrow \mathsf{m} A_p \, .}

\paragraph*{Case b:} 
If we find that $\mathsf{s} \prod_{v \in \tilde{p}'} B_v|\psi\rangle=|\psi\rangle$ for all possible choices of $g_{p'}$ in Eq.~\eqref{eq:app-pm}, then the set of stabilizers remains unchanged. This suggests that the state $|\psi\rangle$ must be an eigenstate of the measurement operator $A_p$ with eigenvalue $\mathsf{m}=\pm1$. Let us assume that the operator $A_p$ can be expressed in terms of the instantaneous stabilizer generators as 
\eq{eq:ap-comm-resolved}{\mathsf{m}A_p = \prod_i g_i^{\mu_i} \prod_j q_j^{\nu_j}\prod_k \mathcal{Z}_k^{\xi_k}}
where the variables $\mu,\nu, \xi\in\{0,1\}$ denote the presence or absence of the corresponding generator in the operator string. Recall from the definitions in Eqs.~\eqref{eq:Qdefinition} and \eqref{eq:Gdefinition} that the individual operators $q$ and $\mathcal{Z}$ are products of Pauli-Z operators whereas the operators $g$ additionally involve plaquette stabilizers $A$. The goal is to find out the value of $\mathsf{m}$ by determining which generators multiply together to give $\mathsf{m}A_p$.

Since the plaquette destabilizer $d_i$ anticommutes only with its partner $g_i$ and commutes with the rest of $g\neq g_i$, we conclude that $\mu_i=1$ if and only if $\{A_p,d_i\}=0$. In principle, we can also detect the presence of vertex stabilizers $q$ by tracking the evolution of the corresponding vertex destabilizers. Unfortunately, the operators that anticommute with only a pair of $B_v$ would in general be linear-depth circuits since they correspond to creating point-like non-Abelian anyons. Instead, we leverage the underlying lattice structure of the stabilizers to find the non-zero values of $\nu$ and $\xi$.

Let us consider the product $A_p\prod_{i:\mu_i=1}g_{i}$ which, according to Eq.~\eqref{eq:ap-comm-resolved}, must be of the form $\mathsf{s'}\prod_{l\in\mathcal{E}}Z_l$ where $\mathsf{s'}=\pm 1$ is an overall sign. For simplicity, let us first consider the case where $\xi_k=0$, for which we do not require non-contractible loop operators to determine the measurement outcome. Then we have
\eq{}{ \eqsp{\mathsf{m} =& \big(A_p\prod_{i:\mu_i=1}g_{i}\big) \  \prod_j (\mathsf{s_{q_j}}B_j)^{\nu_j}\\
&=\big(\mathsf{s'}\prod_{l\in\mathcal{E}}Z_l\big) \prod_j \  (\mathsf{s_{q_j}}B_j)^{\nu_j} = \mathsf{s'}\times \prod_{j:\nu_j=1}\mathsf{s_{q_j}}
}}
where we have explicitly separated the signs $\mathsf{s_{q_j}}$ in the instantaneous vertex stabilizers from their operator $B$, as per the definition in Eq.~\ref{Eq:Stabilizersevolved}. The last equality can hold only if $Z_{l\in\mathcal{E}}$ precisely cancel the $B$ operators for which $\nu$ is nonzero. This is possible if and only if $\nu_j=1$ for vertices that are bounded by edges in $\mathcal{E}$ and zero otherwise. So, we find that the measurement outcome $\mathsf{m}$ is a product of the sign $\mathsf{s'}$ and the parity of the number of vertex defects bounded by edges in $\mathcal{E}$. This parity can be efficiently computed by counting the number of times the path that pairs up all vertex defects (generated using the pyMatching package~\cite{higgott2023pymatching}) crosses the boundary $\mathcal{E}$. The case with a non-contractible loop, i.e. $\xi_k=1$, can be treated analogously. Here, the relevant quantity to consider is the parity of defects that is bounded by $\mathcal{E}$ and the path of the logical $Z$ operator that is parallel to the non-contractible loop present in $\mathcal{E}$.

The updates of logical operators as a result of the measurements are carried out similarly. The $Z$ logical operators commute with both $B_v$ and $A_p$ operators and hence remain unchanged. The logical $X$ operators have non-trivial commutation relations with $A_p$ as a result of Eq.~\eqref{eq:cz-relation}. If, at a certain time step, measurement operators $A_p$ anticommute with the logical (i.e. $\{A_p,\mathcal{X}_L\}|\psi\rangle=0$) in addition to some other stabilizers $g$, then we update the logical $X$ operator as discussed in case a) above. If we find that the measurement \emph{only} anti-commutes with the logical $X$ stabilizer, then it would imply that the state is no longer stabilized by the logical $X$ operator after the measurement. We term it as a logical error since it collapses the logical $X$ eigenstate. 

\subsection{Stabilizer tableaux}

Having identified the form of instantaneous stabilizers, we now present the method used in our simulations for encoding them efficiently. The series of steps discussed in the previous subsection only switches the signs $\mathsf{s_{v}}$ associated with the vertex stabilizers $q_{v}=\mathsf{s_{v}}B_v$ without modifying the form of the operator. Hence, the instantaneous state of the vertex stabilizers can be represented by a length $2L^2$ array whose elements keep track of these signs. 

The remaining plaquette-type stabilizers in $\mathcal{G}$ are modified in a non-trivial manner, which includes multiple $A$ operators with additional Pauli-Z decorations in the given stabilizer. More explicitly, a generator $g_j\in \mathcal{G}$ can be written in terms of its constituents as 
\eq{}{g_j = \mathsf{s_j}\prod_{k=1}^{N_q} (Z_k)^{\mathsf{z_{j,k}}}\prod_{p=1}^{N_p} (A_p)^{\mathsf{a_{j,p}}} }
where $N_q=3L^2$ and $N_p=L^2$ is the total number of qubits and plaquettes in the system, respectively. Here, the products run over all three colors $r,g,b$ but we suppress the color index for brevity. The variables $\mathsf{z_{j,k}}$ and $ \mathsf{a_{j,k}}$ are either $1$ or $0$ depending on whether the associated operator is present or absent respectively and $\mathsf{s_j}=\pm1$ is the overall sign. As discussed in Eq.~\eqref{eq:Gdefinition}, there are $L^2-3=N_p-3$ independent generators. We collect all of these operator representations in a $(N_p-3)\times(N_q+N_p)$ dimensional matrix
\eq{}{\mathsf{G} = \left[\begin{array} {ccc|ccc}
    \mathsf{z_{1,1}} & \dots  & \mathsf{z_{1,N_q}} & \mathsf{a_{1,1}} & \dots  & \mathsf{a_{1,N_p}}\\
    \vdots & \ddots & \vdots&\vdots & \ddots & \vdots\\
    \mathsf{z_{N_p-3,1}} & \dots  & \mathsf{z_{N_p-3,N_q}} & \mathsf{a_{N_p-3,1}} & \dots  & \mathsf{a_{N_p-3,N_p}}
    \end{array} \right]}
where each entry is either $0$ or $1$. The overall sign $\mathsf{s_{j}}$ is stored in a separate length $N_p-3$ array. The state of destabilizers can be analogously tracked using matrix $\mathsf{D}$ where $j^{th}$ row represents the instantaneous operator $d_j$.

A key feature of this representation is that the matrices $\mathsf{G}$ and $\mathsf{D}$ can be updated efficiently. The update after application of $X$ or $Z$ to a qubit has a non-trivial effect on the stabilizer only if the plaquette sharing that qubit is present in its operator string. Hence, such an update only requires scanning through two columns of the matrix to determine the modified stabilizers. The updates are done by adding new $Z$ decorations or changing the sign, which involves updating only $O(1)$ entries per row. The implementation of stabilizer measurement requires determination of the commutation relation between $A_p$ and a row of $\mathsf{G}$ matrix. The operator $A_p$ has non-trivial commutation relations only with 6 nearby plaquettes according to Eq.~\eqref{eq:d4-commute-relation}. Hence, the checks similar to Eq.~\eqref{eq:app-pm} to determine if we are in case (a) or case (b) discussed above require scanning through $6$ columns of the matrix per measurement event and has $O(L^2)$ operation cost. The simulation of deterministic measurement outcome (see case b) has an additional overhead of generating matching graph that pairs all defects. We determine these pairings using pyMatching package that solves this problem in $poly(L)$ time \cite{higgott2023pymatching}. The part of the update rules that depend on algebraic relations between stabilizer operators $A,B$ and Pauli operators are pre-computed at the beginning of the simulation. This includes indices of qubits coupled to a given qubit by $CZ$ gates and indices of vertex operators that reside in the interior of an overlapping plaquette. Since this is still $O(1)$ data per qubit/plaquette, it does not change the overall memory requirement to simulate the given system. The locations of $CZ$ gates in the logical $X$ string are also pre-computed, which incurs a similar cost ($O(L)$ for each of the three $X$-logicals).


\begin{figure}

\subfloat{%
  \includegraphics[width=\columnwidth]{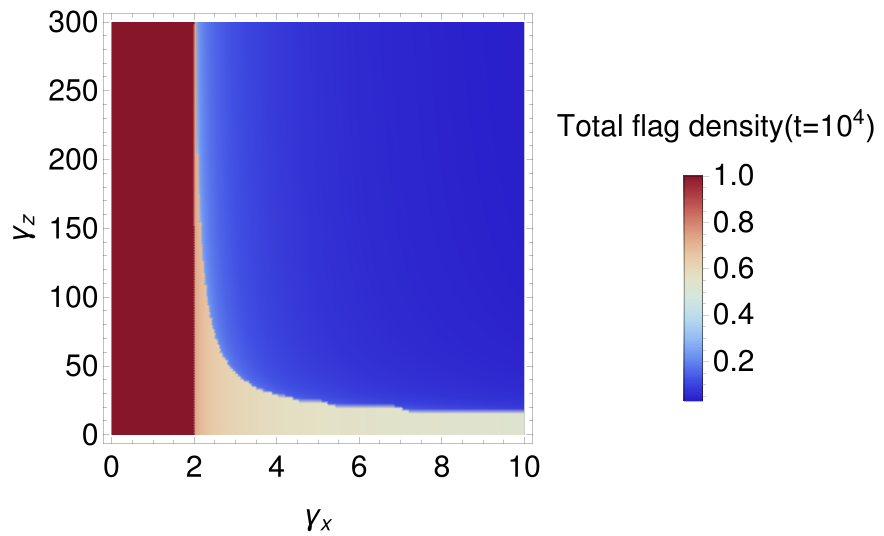}%
}\hfill
\caption{Mean field phase diagram of flag densities in the $D_4$ model: the color is proportional to the total density of flags $(n^f_X+n^f_Z)/2$ evaluated at a long time $t=10^4$. }\label{fig:d4-mf}
\end{figure}

\section{Mean field equations for erasure dynamics}
\label{app:mf}

In this Appendix, we discuss a simple mean field picture of the dynamics of flags in the $D_4$ model. We model the dynamics using a site-approximation where we neglect the correlations amongst flags located on different sites. In particular, multi-site correlation functions $\tr{(\rho n^f_j n^f_k)}$ in state $\rho$ are approximated by the product of their average values $\tr{(\rho \ n^f)}^2$. Here, we additionally assume the densities to be translation invariant. The mean field dynamics is then described in terms of the $X$ flag density $n^f_X$ and the $Z$ flag density $n^f_Z$. 

The density of $X$ flags evolves in time according to 
\eq{eq:meanfield-x}{\frac{d n^f_X}{dt} = \eta (1-n^f_X) - 2\gamma_x n^f_X(1-n^f_X)^2 \, . }
Here, the first term corresponds to the addition of $X$ flags at unoccupied sites after an erasure error. The second term models the contribution from the leaf-correction protocol, where an $X$ flag on a given edge is removed if it is the only flagged edge incident on one of its vertices. We note that the loop moves in the main text were introduced to remove closed loops of flags that are otherwise unaffected by the leaf correction operations. Since the mean field approximation neglects any inter-site correlations, it does not distinguish between leaf and loop configurations. Thus, the correction term in Eq.~\eqref{eq:meanfield-x} is sufficient to remove all flags in this approximation, and we neglect the contributions from loop moves to obtain a simplified description.

The dynamics of $Z$ flags is coupled to that of $X$ flags as they evolve according to 
\eq{eq:meanfield-z}{\eqsp{
\frac{d n^f_Z}{dt} &=  \eta (1-n^f_Z) + 2\gamma_x n^f_X(1-n^f_X)^2 (1-n^f_Z)  \\ &  - 2\gamma_z n^f_Z (1-n^f_Z)^5 (1-n^f_X)^6\, .
}}
Here, the first term corresponds to the addition of a $Z$ flag as a result of an erasure error and the second terms adds a new $Z$ flag to an unoccupied site as a result of the $X$ correction procedure. The leaf correction procedure is accounted for by the last term, where the factor $(1-n^f_X)^6$ restricts the correction to the $Z$ flags that do not have any $X$ flags in their vicinity.

We numerically integrate the mean field Eqs.~\eqref{eq:meanfield-x} and \eqref{eq:meanfield-z} up to some long time $t$ to obtain the mean-field phase diagram shown in Fig.~\ref{fig:d4-mf}. At sufficiently high correction rates, the flags remain in an active state with small densities. In contrast, for small correction rates, they enter an absorbing state. These two phases are separated by an intermediate partially active phase where only $X$ flags remain active. We note a qualitative difference compared to the genuine phase diagram, shown in Fig.~\ref{fig:2dphase}(b): for large values of $\gamma_z$, the fully active phase of the mean field model directly transitions into the absorbing phase without going into the intermediate partially active phase. This is not surprising since the partially active state in the un-approximated model appears because $Z$ flags can not be corrected as long as $X$ flags inside the nearby plaquettes are not removed. However, the $(1-n^f_X)^6$ factor in Eq.~\eqref{eq:meanfield-x} only approximates the averaged number of such plaquettes and hence the correction rate is overestimated by the mean field equations. 


\begin{figure}
\includegraphics[width=\columnwidth]{{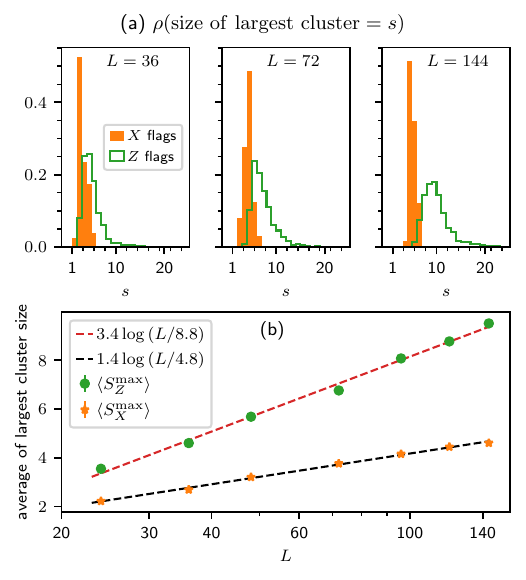}}
\caption{\label{fig:cluster-dist} Connected clusters of flags in active phase: The largest connected cluster of flags is determined in each Monte Carlo trajectory after $500$ sweeps evaluated at a representative point in the active phase $\gamma_x=35,\gamma_z=500$. (top panel) Probability density $\rho(s)$ that the largest cluster in a given Monte Carlo realization is of size $s$ is plotted for various system sizes for both $X$ and $Z$ type flags. The size $s$ of a cluster is defined as the total number of flags in the connected cluster. The data is shown for $1000$ samples. (bottom panel) The average size of the largest cluster $\langle S^{\text{max}}\rangle$ obtained from the histograms is plotted as a function of increasing system size $L$. The error bars represent the standard error in the average value, and dashed lines are obtained by fitting the data points to function $a\log{(L/L_0)}$. }
\end{figure}

\section{Distribution of connected flag-clusters}
\label{app:cluster}

In this Appendix, we present further details of the mixed steady-state $\rho$ stabilized in the active phase of the $D_4$ model. When this state $\rho$ is driven purely by the noiseless evolution, generated by the correction jump operators in $\mathcal{L}_C$ (see Eq.~\eqref{eq:d4corr}), it approaches a ground state of the $D_4$ Hamiltonian Eq.~\eqref{Eq:D4Ham}, while all flags approach the $n^f=0$ state. Specifically, as shown in Fig.~\ref{fig:d4-two-way-fid}, the time it takes for the infidelity $1-F(\rho,|\psi_{D_4}\rangle)$ between a genuine $D_4$ ground state and the recovered state to fall below a fixed constant $\varepsilon$ increases as $\sim \log{L}$ with increasing system size $L$. Note that the local moves of the correction channel remove the connected clusters of flags only from their endpoints. Intuitively, this suggests that the time for the recovery should be controlled by the largest cluster of flags in the original state $\rho$. In Fig.~\ref{fig:cluster-dist}, we analyze the distribution of such largest connected clusters of flags in the Monte Carlo trajectories that sample the active steady-state $\rho$. The average size of the largest cluster scales as $\log{L}$ which qualitatively agrees with the recovery timescale computed using the fidelity.



\newpage
\bibliography{ref.bib}

\end{document}